\documentclass[twocolumn,amsmath,amssymb,showkeys,showpacs]{revtex4}

\usepackage{graphicx}
\usepackage{dcolumn}
\usepackage{bm}
\usepackage{float}
\usepackage{amsfonts}

\begin{document}

\preprint{APS/XYZ}

\title{Fermi surface induced lattice distortion in NbTe$_2$}

\author{Corsin Battaglia}
\affiliation{Institut de Physique, Universit\'e de Neuch\^atel,
CH-2000 Neuch\^atel, Switzerland}%
\email{corsin.battaglia@unine.ch}
\homepage{http://www.unine.ch/phys/spectro}
\author{Herv\'{e} Cercellier}
\affiliation{Institut de Physique, Universit\'e de Neuch\^atel,
CH-2000 Neuch\^atel, Switzerland}%
\author{Florian Clerc}
\affiliation{Institut de Physique, Universit\'e de Neuch\^atel,
CH-2000 Neuch\^atel, Switzerland}%
\author{Laurent Despont}
\affiliation{Institut de Physique, Universit\'e de Neuch\^atel,
CH-2000 Neuch\^atel, Switzerland}%
\author{Michael Gunnar Garnier}
\affiliation{Institut de Physique, Universit\'e de Neuch\^atel,
CH-2000 Neuch\^atel, Switzerland}%
\author{Christian Koitzsch}
\altaffiliation[Now at ]{ABB, CH-5600 Lenzburg, Switzerland}
\affiliation{Institut de Physique, Universit\'e de Neuch\^atel,
CH-2000 Neuch\^atel, Switzerland}%
\author{Helmuth Berger}
\affiliation{Institute of Physics of Complex Matter, EPFL, CH-1015
Lausanne, Switzerland}
\author{L\'{a}szl\'{o} Forr\'{o}}
\affiliation{Institute of Physics of Complex Matter, EPFL, CH-1015
Lausanne, Switzerland}
\author{Claudia Ambrosch-Draxl}
\affiliation{Institut f\"{u}r Physik,
Karl-Franzens-Universit\"{a}t Graz, A-8010 Graz, Austria}
\author{Philipp Aebi}
\affiliation{Institut de Physique, Universit\'e de Neuch\^atel,
CH-2000 Neuch\^atel, Switzerland}%

\date{\today}

\begin{abstract}
The origin of the monoclinic distortion and domain formation in
the quasi two-dimensional layer compound NbTe$_2$ is investigated.
Angle-resolved photoemission shows that the Fermi surface is
pseudogapped over large portions of the Brillouin zone. \textit{Ab
initio} calculation of the electron and phonon bandstructure as
well as the static RPA susceptibility lead us to conclude that
Fermi surface nesting and electron-phonon coupling play a key role
in the lowering of the crystal symmetry and in the formation of
the charge density wave phase.
\end{abstract}


\maketitle

\section{\label{sec:level1}Introduction}

NbTe$_2$ belongs to the category of layered transition metal
dichalcogenides (TMDC) known for their quasi two-dimensional (2D)
properties. Due to the reduced dimensionality, free charge
carriers and phonons are coupling in an unique fashion, leading to
the formation of charge density
waves (CDW) and  superconductivity.\\
The competition between these two electronic groundstates is
especially interesting in view of the anomalous properties of
another class of strongly anisotropic layered materials, the high
T$_c$ cuprate superconductors. Attempts to explain the complexity
of electronic and magnetic properties observed in these compounds
are based on the subtle balancing of competing interactions
producing superconducting pairing, spin and charge ordering. For
NbTe$_2$, magnetic degrees of freedom are unlikely to play an
important role. Thus, in principle NbTe$_2$ allows to isolate the
effects associated with density wave instabilities and
superconductivity.\\
An explanation for the CDW transition in 2D materials is derived
from the theory for the Peierls instability \cite{Peierls55} in 1D
metals. A system of conduction electrons may under suitable
conditions become unstable with respect to a spatially modulated
perturbation with wavevector $\mathbf{q}$, such as a static
periodic lattice distortion. Kohn \cite{Kohn59} has shown that
such soft phonon modes may result from the screening of lattice
vibrations by conduction electrons. According to linear response
theory, the quality of the screening by the electrons is measured
by the static generalized susceptibility with Fourier component
$\chi(\mathbf{q})$. Instability sets in when this quantity
diverges. This happens under favorable nesting conditions for
which large portions of the Fermi surface can be connected or
nested by a single $\mathbf{q}$ vector. Even when the system is
not truly one-dimensional, nesting may become important, if the
Fermi surface consists of flat parallel sheets. However 2D systems
often remain metallic, since the opening of the gap removes only
parts of the Fermi surface.\\ \begin{figure}[b] \centering
\includegraphics{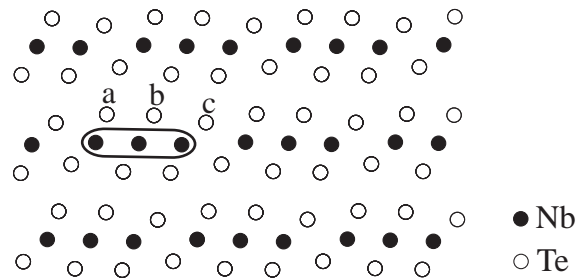}%
\caption{\label{fig:struct1}Side view of the monoclinic structure
of NbTe$_2$ (projection onto the (010) plane within the monoclinic
space group). Each layer of NbTe$_2$ consists of a Te-Nb-Te
sandwich. The Nb atoms (filled circles) are displaced within the
plane and form 'trimers', whereas the Te atoms (empty circles)
exhibit an out-of-plane buckling (a,b,c). Successive Te-Nb-Te
sandwiches are shifted within the plane. The stacking sequence is
repeated after 3 layers.}
\end{figure}
The distortion already observed at room temperature in NbTe$_2$
and the isostructural TaTe$_2$ suggests the action of a
single-axis CDW. The structure \cite{Brown66} is a monoclinically
deformed version of the trigonal \textit{1T} polytype, in which
the transition metal sits in octahedrally coordinated sites
between the chalcogen atoms (see Appendix \ref{app} for structural
details). The metal atoms are displaced from the center of the
coordination unit and the chalcogen layers form zig-zag chains to
accomodate these shifts (see Fig. \ref{fig:struct1}). After
cooling of heat pulsed crystals to room temperature, transmission
electron diffraction experiments revealed a second, seemingly
unrelated triple-axis CDW state with a
$(\sqrt{19}\times\sqrt{19})$ signature
\cite{Landuyt74,Landuyt75,Wilson78}, commensurate at room
temperature, but readily rendered incommensurate just above.\\
NbTe$_2$ and TaTe$_2$ are semimetals \cite{Wilson69}. The
resistivity decreases monotonically with decreasing temperature
\cite{Brixner62,Nagata93,Vernes98}. A drop in resistivity of
NbTe$_2$ in the range $0.5-0.74$ K marks the transition into the
superconducting phase \cite{Maaren67}. Superconductivity is absent
in TaTe$_2$ \cite{Kidron67}.\\
After a discussion of the experimental and theoretical methods, we
will investigate the interplay between electronic and structural
properties of NbTe$_2$. Our low energy electron diffraction (LEED)
measurements will allow an alternative parametrization of the
structure proposed by X-ray diffraction measurements, which is
more appropriate in the discussion of electron dynamics. Numerous
studies of the electronic properties have been carried out for
disulfides
\cite{Pillo99b,Pillo01,Pillo02,Aebi01,Bovet03,Bovet04,Clerc04,Clerc04b,
Perfetti05} and diselenides
\cite{Aebi01,Horiba02,Perfetti03,Bovet04,Colonna05}. Direct
measurements of the Fermi surface topology of NbTe$_2$ and
TaTe$_2$ have never been reported. In order to shed light on the
origin of the CDW phase and the domain formation observed in these
compounds, we have measured the Fermi surface via
full-hemispherical angle-resolved photoelectron spectroscopy
(ARPES). Scanning tunneling spectropscopy (STS) experiments
complement the ARPES data. We suggest that the distortion and the
accompanying domain formation is intimately related to the Fermi
surface topology. Quantitative assessments of its nesting
tendencies are obtained from first principle bandstructure
calculations. A computation of the vibrational spectrum and a soft
mode analysis support our conclusions. \\
In the following, we concentrate on NbTe$_2$. Similarities and
differences between NbTe$_2$ and TaTe$_2$, especially the absence
of superconductivity in TaTe$_2$, will be addressed at the end.

\section{Experiment and Calculation}
\begin{figure}
\includegraphics{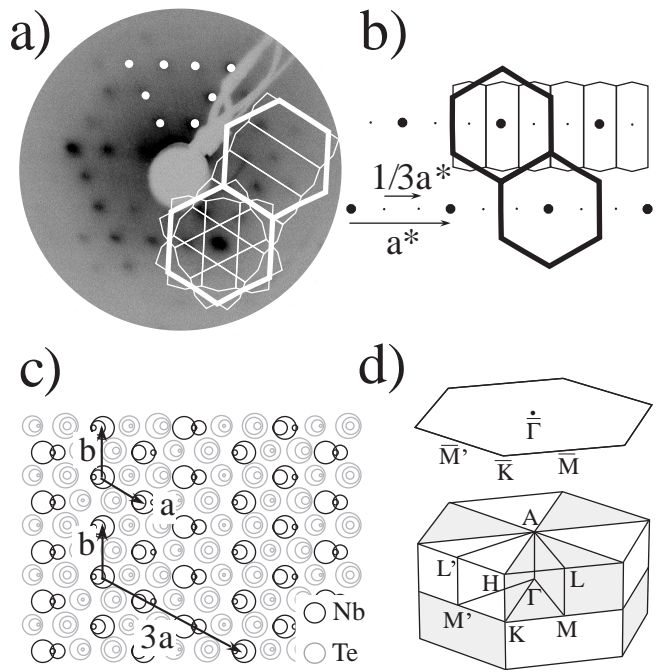}%
\caption{\label{fig:LEED}a) LEED pattern (75 eV) of NbTe$_2$ at
room temperature. b) Schematic illustration of a single domain $(3
\times 1)$ superstructure as observed in reciprocal space. The fat
dots indicate the $\Gamma$ points of the undistorted trigonal
lattice. Due to the $(3\times 1)$ distortion in real space, two
additional spots appear in between these main reflections
represented by the small dots. Their spacing is $\frac{1}{3} a^*$.
The bold hexagons outline the surface Brillouin zone of the
trigonal $(1\times 1)$ structure. The thin compressed hexagons
show the surface Brillouin zone of the $(3\times 1)$ superlattice.
c) Projection onto the basal plane of the first three Te-Nb-Te
sandwiches of the monoclinic structure. The size of the circles is
proportional to the $z$-coordinate. The lattice vectors of the
$(1\times 1)$ and $(3 \times 1)$ cells are indicated. d) Bulk and
surface Brillouin zone for \textit{1T}-NbTe$_2$ with high symmetry
points.}
\end{figure} Full hemispherical ARPES
experiments were performed in a modified Vacuum Generator ESCALAB
Mark II spectrometer with a residual gas pressure of $2\times
10^{-11}$ mbar equipped with a Mg K$_\alpha$ ($h \nu= 1253.6$ eV)
X-ray anode, a monochromatized He discharge lamp providing He
I$\alpha$ ($h\nu=21.2$ eV) radiation \cite{Pillo98}, and a three
channeltron hemispherical electrostatic analyzer kept fixed in
space during measurements. The sample is mounted on a manipulator
with two motorized and computer controlled rotational axes and may
be cooled via a closed cycle refrigerator. Energy resolution is 50
meV, the combined angular resolution of sample manipulator and
analyzer is approximately $1^o$. Surface cleanness and quality
before and after ARPES measurements was monitored by X-ray
photoelectron spectroscopy (XPS) and checked with low energy
electron diffraction (LEED) respectively. Orientation of the
sample was achieved by X-ray photoelectron diffraction (XPD)
\cite{Aebi97,Fasel02}.\\ Scanning tunneling microscopy (STM) and
spectroscopy (STS) experiments were carried out with Pt/Ir tips
using an Omicron LT-STM in a separate UHV system with a base
pressure of $3\times 10^{-11}$ mbar. Pure NbTe$_2$ samples were
prepared by the standard flux growing techniques. Sample cleavage
was
carried out in UHV using adhesive tape.\\
First principle calculations were performed in the framework of
density functional theory (DFT) using the full potential augmented
plane wave plus local orbitals (APW+lo) method in conjunction with
the generalized gradient approximation (GGA) in the
parametrization of Perdew, Burke and Ernzerhof \cite{Perdew96} as
implemented in the WIEN2k software package \cite{wien2k} as well
as the ABINIT code \cite{Gonze02,abinit} using the local density
approximation (LDA) and relativistic separable dual-space Gaussian
pseudopotentials \cite{Hartwigsen98} for both Nb and Te, taking
into account the Nb
semicore states.\\
A recent extension to WIEN2k based on the OPTICS package allows
the computation of the frequency dependent random phase
approximation (RPA) susceptibility \cite{Ambrosch04}. The phonon
dispersion is computed with the help of the linear response or
density functional perturbation theory (DFPT) capabilities of
ABINIT \cite{Gonze97,Gonze97b}. Computational details are given in
\cite{DFTdetails}.

\section{Results and Discussion}

\subsection{$(3\times 1)$ surface superstructure}

Transmission electron microscope (TEM) images of NbTe$_2$ show a
complicated domain structure (see \cite{Wilson69}). Although the
crystal structure within the domains is known, insights into the
domain structure can be obtained from LEED experiments. Figure
\ref{fig:LEED}a) presents a LEED measurement taken with electrons
accelerated to a kinetic energy of $75.0$ eV. An interpretation
based on the monoclinic reciprocal lattice is not straightforward.
The experimental pattern is most easily understood by considering
the undistorted trigonal parent structure. Since LEED patterns
exhibit Bragg reflections of the 2D surface lattice, the surface
Brillouin zone borders are superimposed. The bold hexagons
correspond to the surface Brillouin zone of the trigonal $(1
\times 1)$ structure. Due to the monoclinic distortion of the
lattice, two additional spots in between the main reflections are
visible, which can be understood in terms of a $(3 \times 1)$
superstructure. A schematic illustration is given in Fig.
\ref{fig:LEED}b). The presence of this superstructure results in a
new surface Brillouin zone shown in Fig. \ref{fig:LEED}b). The
LEED pattern results from the superposition of three orientational
variants, rotated by 120$^o$ with respect to eachother, of such
surface superstructures.  Note that a $(3 \times 3)$
superstructure would result in additional spots occurring in the
center of the triangle outlined by the white dots in Fig.
\ref{fig:LEED}a). A close inspection of the monoclinic structure
in Fig. \ref{fig:LEED}c) reveals a $( 3 \times 1)$ surface unit
cell, confirming the experimental finding. The presence of
additional spots were already reported earlier \cite{Cukjati02b}
and correctly interpreted as a superposition of patterns from
three domains. However, the appearance of these superspots has not
been recognized as a $(3 \times 1)$ surface superstructure. For
completeness, we note that the bulk structure exhibits a $(3\times
1\times 3)$ supercell structure, since successive layers are
shifted within the plane (see Fig. \ref{fig:struct1}).\\
The diffuse nature of the LEED pattern has been explained in terms
of thermal disorder \cite{Cukjati02b}. Our LEED measurements at
low temperature remain diffuse. The presence of a fine domain
structure might be responsible for the broadening of the
reflections. The average domain size is obtained by comparison of
the width of the reflections with the inter-reflection distance.
From the measurement in Fig. \ref{fig:LEED}a) with an average peak
width of 0.2 \AA$^{-1}$, we estimate the average domain size to be
of the order of 32 \AA, which agrees quite well with the domain
size from a sample from the same batch observed by STM (see below).\\
LEED is not able to distinguish between three or six orientational
variants of the $(3 \times 1)$ superstructure. Whereas LEED probes
the periodicity of the surface, XPD indicates the symmetry of the
local environment of the emitting atom. The Te 3$d_{5/2}$ XPD
diffractogram (not shown) exhibits a three-fold symmetry, which
clearly shows that only three and not six orientational variants
are present, since the presence of domains rotated by 180$^o$
would result in a six-fold symmetry. We observed further that XPD
diffractograms from different cleavage planes were rotated by
180$^o$ with respect to eachother, retaining however their
three-fold symmetry, indicating a
change in the stacking sequence between successive Te-Nb-Te layers.\\
The presence of the $(3 \times 1)$ superstructure implies a
reconstruction of the Brillouin zone. In ARPES experiments, we may
thus expect to observe the opening of a gap at the new Brillouin
zone border accompanied by a backfolding of bands.

\subsection{Fermi surface topology}
\begin{figure}
\includegraphics{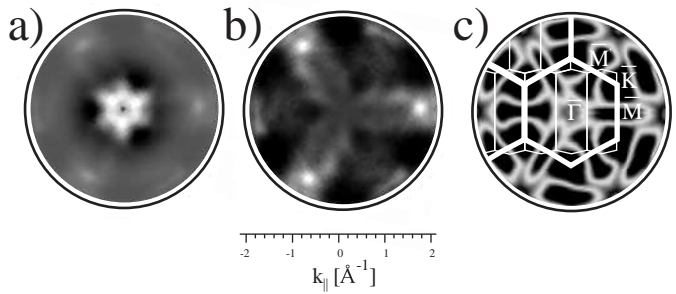}%
\caption{\label{fig:FSM} a) Angular distribution of electrons from
$E_F$ mapped as a function of $\mathbf{k}_{||}$ at room
temperature, white and black correspond to high and low intensity,
respectively. b) Symmetrized and flattened data of a). c)
Theoretical APW+lo Fermi surface contours for the undistorted
trigonal NbTe$_2$ in the free electron final state approximation
\cite{Aebi01} with photon energy $h\nu=21.2$ eV, workfunction
$\phi=4$ eV and an assumed inner potential $V_0=13$ eV. Brillouin
zones of the $(1\times 1)$ and one particular orientation of the
$(3\times 1)$ lattice are superimposed. }
\end{figure}
\begin{figure*}
\includegraphics{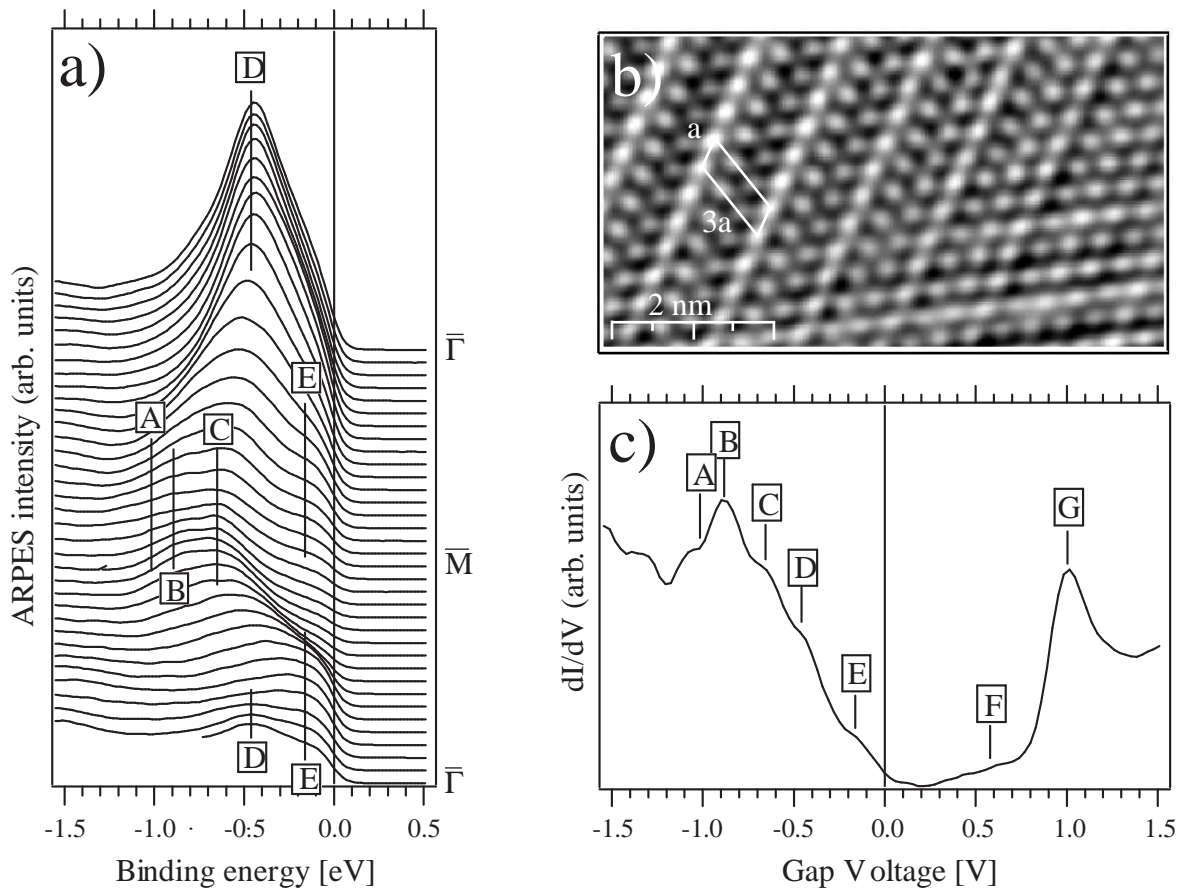}%
\caption{\label{fig:STS} Comparison between ARPES and STS data. a)
Room temperature ARPES spectra along
$\bar{\Gamma}\mathrm{\bar{M}}$. b) STM topography of the $(3\times
1)$ phase of NbTe$_2$ at 77 K (V=3 mV, I=2.9 nA). c) STS spectrum
at 77 K measured with a lock-in amplifier, modulation frequency 1
kHz, modulation amplitude 30 mV. The sample-tip distance was
chosen in order to have a tunneling current of 0.8 nA for a bias
voltage of -1.5 V.}
\end{figure*}

Figure \ref{fig:FSM} presents Fermi surface maps (FSM) of NbTe$_2$
measured at room temperature, i.e. the intensity distribution for
electrons from the Fermi level ($E_F$). These maps bear in many
respects a close resemblance to the FSM's of isopolytopic TaS$_2$
and TaSe$_2$ \cite{Bovet04}, although in these materials, the CDW
distortion is of the $(\sqrt{13}\times\sqrt{13})$ type. Figure
\ref{fig:FSM}a) gives experimental raw data without any further
treatment. Near normal emission, high intensity is measured which
falls off rather quickly towards larger polar angles. A similar
behavior is observed for \textit{1T}-TaS$_2$ and
\textit{1T}-TaSe$_2$ and has been attributed to the $d_{z^2}$
character of the transition metal band. A normalization of the FSM
by the mean intensity for each polar emission angle as shown in
Fig. \ref{fig:FSM}b) eliminates this dependence and allows to
reveal weaker off-normal features. Centered circular
features are then suppressed. \\
The washed out character of the experimental FSM contours is
another common feature of the \textit{1T} family. Since the width
in $\mathbf{k}_{||}$ of the bands is independent of temperature,
we discharge thermally populated phonons as the origin of the
broadening. From the monoclinic distortion it is expected, that
the $d_{z^2}$ band, which is mainly responsible for the spectral
weight observed at the Fermi energy, splits into several subbands.
This results in a larger width of the observed band in energy and
consequently also in momentum. Secondly, in the presence of the
domain structure, the coherence length of Bloch electrons must be
of the order of the average domain size, since electrons get
scattered at the domain boundaries. As in LEED, this should lead
to a broadening of the crystal momentum of about 0.2 \AA$^{-1}$,
consistent with the broadening observed in the experimental FSM.
 \\
Great care is required in determining the Fermi surface crossings
from the experimental data. Due to the weak dispersion and
$\mathbf{k}$-dependent photoemission matrix elements, which lead
to intensity variations which have nothing to do with Fermi
crossings, an unambiguous extraction of the Fermi surface needs
additional information. We have measured the energy dispersion of
the NbTe$_2$ $d_{z^2}$ band along $\bar{\Gamma}\mathrm{\bar{M}}$
(Fig. \ref{fig:STS} a)), $\bar{\Gamma}\mathrm{\bar{M}'}$ and
$\bar{\Gamma}\mathrm{\bar{K}}$ at room temperature and at T$<$20 K
(not shown). The band topology is not affected by this change in
temperature indicating the absence of a phase transition in this
temperature range.
In order to determine the location of Fermi crossings, we applied
the symmetrization method described in Ref. \cite{Mesot01}. No
quasiparticle crossing has been found for any of the measured
energy dispersion curves. Thus, strictly speaking, the maps in
Fig. \ref{fig:FSM} are not Fermi surfaces. The observed intensity
originates from bands which come close to the Fermi level, but
must not be associated with quasiparticle crossings, but rather
spectral weight which leaks across the Fermi level. The signature
of such a pseudogapped Fermi surface is also observed in TaS$_2$
and TaSe$_2$ \cite{Bovet04} as well as in high T$_c$
superconductors \cite{Damascelli03} and has remained a
controversial topic. \\
Comparison of the experimental data with
the theoretical DFT Fermi surface map shown in Fig.
\ref{fig:FSM}c) for the undistorted trigonal structure shows that
the symmetry of the undistorted Fermi surface is clearly dominant.
In contrast to the superspots in LEED data, no clear evidence for
the superstructure in the form of additional backfolded features
is apparent. Voit \textit{et al.} \cite{Voit00} have argued that
in the presence of a weak superimposed periodic potential, the
spectral weight remains on the unperturbed bands.  An additional
potential due to the $(3 \times 1)$ superstructure manifests
itself through the opening of small gaps localized at the new
Brillouin zone borders.  In the presence of three domains, details
of such a gap structure are buried below the bands of the other
two domains, which do not experience the potential in this
direction, and unfortunately it is impossible
to deconvolute the contributions from the various domains.\\
It should be noted that DFT for the undistorted compound predicts
a truly metallic Fermi surface, whereas our experimental ARPES
data of the distorted structure does not display any quasiparticle
crossings. This indicates that the transition from the high
symmetry structure to the monoclinic structure is driven by a gain
in electronic energy all over the Brillouin zone. The removal of
the entire Fermi surface is not consistent with an explanation
based solely on a 2D Peierls
scenario.\\
Additional information on the occupied and empty state electronic
structure of NbTe$_2$ is obtained via STM. The chain like
structure observed by STM within the domains (Fig. \ref{fig:STS}
b)) is a consequence of the anisotropy of the $(3\times 1)$
lattice and indicates a possible nesting scenario as the origin of
the distortion (see later). For comparison between tunneling and
photoemission spectra, we have pasted the peak positions
determined from the STS spectrum Fig. \ref{fig:STS}c) onto the
ARPES spectra Fig. \ref{fig:STS}a). Although it is difficult to
tell, which part of $k$-space is sampled by STS, features A, B, C,
D and E can be clearly identified with their counterpart in the
ARPES data. As already concluded from ARPES data, the STS spectrum
does not exhibit a clear gap at the Fermi level. In the
pseudogapped region between feature E and G, several small peak
shoulders such as F are observed, indicating a finite density of
states. The theoretical bandstructure discussed in the next
section will allow further
interpretation of these data.\\
The absence of any clear quasiparticle crossing stands in
contradiction with the metallic character of the resistivity
\textit{vs} temperature curves \cite{Nagata93} and can, as
discussed above, not be explained by only taking into account the
Peierls scenario. Electron-phonon coupling is expected to be
relatively strong for the tellurides \cite{Wilson78}, consequently
polaronic effects may play a role. Polarons recently received
increased attention for the interpretation of anomalously broad
ARPES features \cite{Dessau99,Perfetti01,Perfetti02}. Within the
Fermi-liquid picture, ARPES peaks are attributed to quasiparticle
excitations, whose lifetime increases when approaching the Fermi
level. This procedure is well justified by recent theoretical
calculations of the spectral function for the spinless Holstein
model in the weak coupling regime \cite{Hohenadler05,Sykora05}. In
contrast in the strong coupling regime, the coherent quasiparticle
band flattens considerably and possesses exponentially small
spectral weight \cite{Hohenadler05,Sykora05}. Most spectral weight
is transferred to a broad incoherent part on the high energy side.
Whereas the center of mass or first moment of the spectral
function remains unaffected by the electron-phonon interaction at
low band-fillings, it experiences a rigid shift proportional to
the polaron binding energy at higher fillings
\cite{Kornilovitch02}. A polaronic scenario would thus allow to
explain the apparent absence of quasiparticle crossings as well as
the broadened line shape in the experimental ARPES spectra of
NbTe$_2$. However, an ideal experimental model system suited for
ARPES measurements, for which the theoretical predictions for
polaronic spectral signatures can be verified, still has to be
found.

\subsection{Electronic bandstructure}
\begin{figure}
\includegraphics{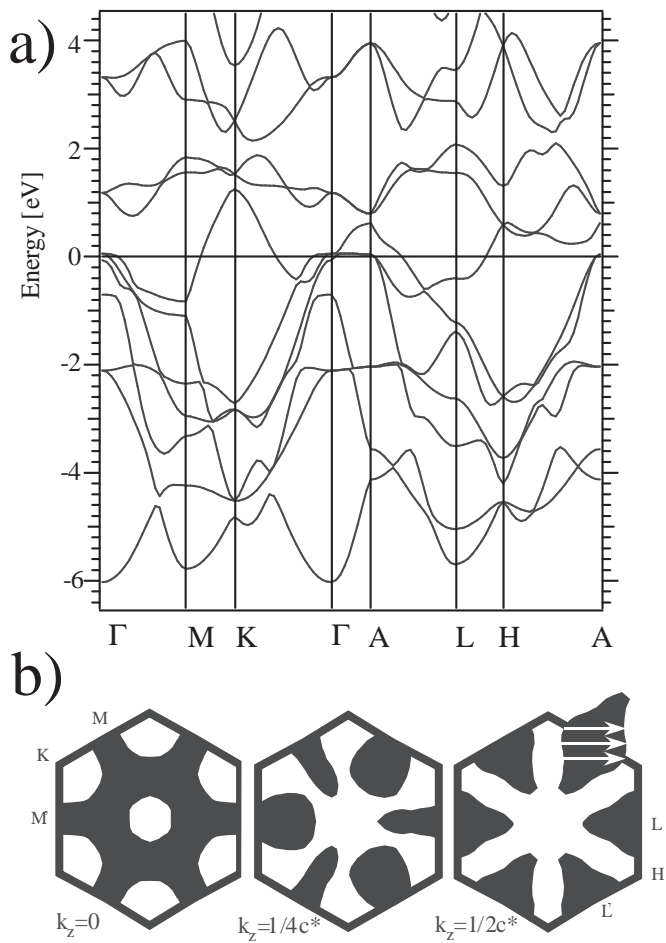}%
\caption{\label{fig:eBS} a) APW+lo bandstructure along high
symmetry directions for \textit{1T}-NbTe$_2$. b) Cuts through the
Fermi surface at $k_z=0$, $k_z=c^*/4$, $k_z=c^*/2$. Nesting
vectors of length $q=1/3 a^*$ are indicated.}
\end{figure}

Comparing NbTe$_2$ and TaTe$_2$ to their homologues in the sulfide
and selenide family, such as TaS$_2$, TaSe$_2$, NbS$_2$ and
NbSe$_2$, it is tempting to interpret the monoclinic distortion in
terms of a CDW phase. The existence of such a phase requires
electron-phonon communication in the undistorted compound.
Trigonal \textit{1T}-NbTe$_2$ is not available for experiment. In
order to make some progress, we investigated the electronic
structure of the undistorted crystal in the framework of DFT (see
Appendix \ref{app} for structural details). \\
In Fig. \ref{fig:eBS}a) we show the bandstructure of
\textit{1T}-NbTe$_2$ along high symmetry directions obtained with
the APW+lo basis set. The positions of the high symmetry points in
the Brillouin zone are indicated in Fig. \ref{fig:LEED}d). The
overall agreement with the bandstructure obtained using the
pseudopotential method (not shown) is very good. Slight variations
can be attributed to the different exchange-correlation functional
and
convergence parameters.\\
From a simple ionic picture, one would expect the Nb$^{4+}$ ions
to have only one remaining $d$ electron, resulting in a
half-filled band crossing the Fermi level, while the six Te $5p$
bands from the two Te atoms are fully occupied. Obviously this
picture neglects all other bonding effects, since an appreciable
admixture of Nb $4d$ states exists in the Te $p$ bands, indicating
covalent interactions and less ionic bonding. Integration of the
partial charges inside the muffin tin sphere of the respective
atom allows to identify the character of the different bands and
correctly reproduces this coarse prediction. The first four empty
bands above
the Fermi level are the remaining four Nb $4d$ bands.\\
Due to the octahedral coordination of the Nb atoms between the Te
atoms and the resulting crystal field, the five Nb $d$ bands are
split into a lower triplet of $t_{2g}$ states and an upper doublet
of $e_g$ states separated by a small gap at 2 eV binding energy.
The $\sigma$ bonding $e_g$ orbitals have higher energies because
they interact strongly with the neighboring Te atoms. The orbital
degeneracy of the octahedral $t_{2g}$ manifold is reduced in a
Jahn-Teller like fashion by a trigonal elongation of the Te
octahedra along the $c$ axis.\\
The P$\bar3$m1 space group contains an unique $z$ axis
perpendicular to the layers of the crystal. Further insights can
be obtained by dividing the $d$ orbitals into (a) $d_{z^2}$
(out-of-plane orientation), (b) $d_{x^2-y^2}$, $d_{xy}$ (in-plane
orientation) and (c) $d_{xz}$, $d_{yz}$. The half-filled,
lowest-lying $t_{2g}$ band, which crosses the Fermi level, has
mainly Nb $d_{z^2}$ character, whereas the remaining two bands of
the $t_{2g}$ manifold (first two unoccupied bands) exhibit
dominant Nb $d_{x^2-y^2}$ and $d_{xy}$ character. While these
in-plane $d$ orbitals do not interact strongly with the Te $p$
orbitals, the $d_{z^2}$ orbitals, due to their orientation towards
the Te layers, are more strongly hybridized with the Te orbitals
especially around the Brillouin zone center. This orbital
resonance might be at the origin of the buckling of the Te atoms.
Whereas the formation of 'trimers' by the Nb atoms is consistent
with a Peierls scenario, the buckling of the Te layer rather
points towards a band Jahn-Teller distortion \cite{Weitering96}.
However, inspection of Fig. \ref{fig:struct1} shows, that the Te
atoms of type c which fall in between the Nb 'trimers' are shifted
towards the Nb layer, whereas the other two Te atoms, labelled a
and b, remain approximately at the original distance. Since the
'trimerization' of the Nb atoms reduces the overlap between the Te
atoms of type c  \ref{fig:struct1} and its neighboring Nb atoms,
these Te atoms approach the Nb layer to recover the overlap.
\\
We now compare the theoretical bandstructure for the undistorted
structure with ARPES and STS data in Fig. \ref{fig:STS}  from the
distorted crystal. We identify feature G on the unoccupied side of
the STS spectrum with the $t_{2g}$ doublet which disperses around
1 eV. These empty bands appear to be only weakly affected by the
reduction of the symmetry towards the monoclinic space group.
 Due to the hybridization of the $d_{z^2}$ band with the Te $p$
bands and the modification of the bandstructure induced by the
distortion to the monoclinic structure, an identification of the
features in the occupied part of the spectrum is not so
straightforward. The theoretical dispersion along $\Gamma$M is
qualitatively in agreement with the ARPES data, where most bands
have the tendency to disperse towards higher binding energies when
going from $\bar{\Gamma}$ to $\mathrm{\bar{M}}$. However, whereas
the Nb $d_{z^2}$ band clearly crosses the Fermi level in the
theoretical bandstructure, no such crossing is observed in our
experimental results. This indicates that the distortion to the
monoclinic structure has profound effects on the Nb $d_{z^2}$
band.
\\
We also computed the bandstructure of NbTe$_2$ in the monoclinic
structure (not shown). The distortion reduces the density of
states at the Fermi level approximately by a factor of 2. However,
the pseudogap observed above the Fermi level in the STS spectrum
(Fig. \ref{fig:STS}) is not reproduced by theory, since a
considerable amount of states is located in this region. This
discrepancy is possibly caused by polaronic effects as discussed
above. Furthermore localization of the electrons decreases the
bandwidth $W$ of the $d_{z^2}$ band and consequently the crucial
parameter $W/U$ with $U$ the on-site Coulomb repulsion energy.
This opens up the possibility of increased correlation effects.
Mott scenarios have been suggested for \textit{1T}-TaS$_2$
\cite{Fazekas79} and \textit{1T}-TaSe$_2$ \cite{Perfetti03}.
However, instead of trying to explain the discrepancy between
experiment and theory with the insufficient treatment of
correlation effects within LDA or GGA, we wish to point out that
the disagreement between theory and experiment may be caused by
the theoretical treatment of the distorted structure within DFT or
more generally within the Bloch theory of periodic crystals. Since
the monoclinic cell contains 18 inequivalent atom positions, we
obtain six times more bands than for the bandstructure calculated
for the trigonal unit cell. The huge amount of bands renders the
comparison with the experimental data difficult. In several
studies on distorted compounds, we noticed that the spectral
weight observed in ARPES experiments tends to exhibit the symmetry
of the undistorted structure and that backfolded features carry
generally only a small amount of spectral weight. The rigorous
backfolding of bands within the theoretical description does not
appear to apply to real systems. A first step towards a
theoretical framework taking into account this observation has
been taken by Voit \textit{et al.} \cite{Voit00}, who weights the
eigenvalues of the distorted structure obtained from a simple
tight-binding model by the projection of the corresponding
eigenvectors onto the eigenvectors of the undistorted structure.
For comparison between theoretical and experimental data, it would
be clearly desirable to implement an equivalent scheme into
\textit{ab initio} codes.
\\
When two portions of the Fermi surface are flat and parallel,
nesting occurs, and the susceptibility diverges logarithmically.
In the isotropic electron gas, favorable nesting conditions are
only encountered in one dimension, where the Fermi surface
consists of two points. However, materials with an anisotropic
Fermi surfaces may exhibit regions where scattering becomes more
singular than in the isotropic electron gas. The presence of the
$(3\times 1)$ superstructure in the monoclinically deformed
structure would then imply a nesting vector
$\mathbf{q}=\frac{1}{3} \mathbf{a^*}$. Figure \ref{fig:eBS}b)
presents three horizontal cross sections through the Fermi surface
of \textit{1T}-NbTe$_2$ for $k_z=|\mathbf{k}_\bot|=0,
\frac{1}{4}c^*$ and $\frac{1}{2}c^*$. The black areas indicate
occupied, the white areas unoccupied states of the Nb $d_{z^2}$
band. Based on a Fermi surface obtained by extrapolation from the
calculated results for the \textit{1T} sulphides and selenides,
Wilson \cite{Wilson78} already noted that a wave vector of
$\mathbf{q}=\frac{1}{3}\mathbf{a^*}$ is to large for nesting
across M and M'. However, since NbTe$_2$ is not an ideal 2D
crystal, nesting becomes possible across L and L' as shown in Fig.
\ref{fig:eBS}b). The arrows indicate one family of the three
symmetry-related experimental Fermi surface nesting vectors
$\mathbf{q}=\frac{1}{3} \mathbf{a^*}$. Furthermore the nested
areas are relatively large encouraging a Fermi surface nesting
scenario. Obviously, the contributions from different nesting
vectors to the static susceptibility are hard to estimate on the
basis of Fermi surface cross sections. In order to obtain
quantitative confirmation for the $\mathbf{q}=\frac{1}{3}
\mathbf{a^*}$ nesting vector, we integrated the DFT bandstructure
to obtain the susceptibility for all vectors within the Brillouin
zone.
\subsection{RPA susceptibility}

\begin{figure}
\includegraphics{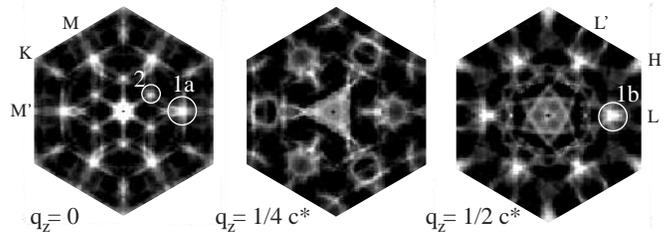}
\caption{\label{fig:sus} RPA susceptibility in the first Brillouin
zone at $q_z=0$ ($\Gamma$MK plane), $\frac{1}{4}c^*$ and
$\frac{1}{2}c^*$ (ALH plane) obtained by integration of the APW+lo
Fermi surface of Fig. \ref{fig:eBS}b). The peaks marked 1a and 1b
occur exactly at $q_x=\frac{1}{3}a^*$. Feature 2 is close to the
nesting vector of the $\sqrt{19}\times\sqrt{19}$ CDW phase.}
\end{figure}

For the computation of the static susceptibility, the following
expression has been used \cite{Koitzsch04}:
\begin{equation}
\chi(\mathbf{q})=\sum_{n',n,\mathbf{k}}
\delta(\epsilon_{n',\mathbf{k}+\mathbf{q}}-\epsilon_{n,\mathbf{k}}).
\end{equation}
The Dirac $\delta$ gives a contribution of either 1 or 0 depending
on whether $\mathbf{q}$ is a nesting vector or not. Matrix
elements are neglected, thus all electron-hole pairs are treated
on an equal basis.
\\
The results of our calculation are presented as linear gray scale
plots in Fig. \ref{fig:sus} with white indicating a large response
of the electron system. Strong nesting is present for small, but
non-vanishing $\mathbf{q}$ vectors. These contributions are due to
intraband contributions from weakly dispersing bands and can be
reduced by choosing a smaller energy window.  \\
Highly interesting is the peak at $\mathbf{q}=\frac{1}{3}
\mathbf{a^*}$ along the $\Gamma$M and $\Gamma$M' directions in
Fig. \ref{fig:sus} (feature 1a). We associate this peak with a
nesting vector leading to the $(3\times 1)$ superstructure
observed by LEED. Thus, the electronic structure of trigonal
NbTe$_2$ appears unstable with respect to a potential with
wavevector $\mathbf{q}=\frac{1}{3} \mathbf{a^*}$. The peak is not
confined to the $\Gamma$MK plane, but is smeared out along $q_z$,
thus allowing an out-of-plane component for the nesting vector. A
second maximum (feature 1b) with the same in-plane coordinates is
seen along the AL and AL' direction. We wish to draw attention to
the fact that the nesting vector sketched in the ALH plane of Fig.
\ref{fig:eBS}b) corresponds to feature 1a, which lies in the basal
plane, whereas feature 1b connects parts of the Fermi surface with
different $k_z$. Our calculation however does not reproduce a
single peak at $\mathbf{q}=\frac{1}{3}
\mathbf{a^*}+\frac{1}{3}\mathbf{c^*}$, which would be required to
explain the occurrence of the $(3\times 1 \times 3)$
superstructure. At present, it is unclear, if the shift between
successive layers is directly induced by the nesting scenario or a
consequence of it. According to the calculation, nesting takes
place in the Nb $d_{z^2}$ band and since the Nb atoms are screened
by the surrounding Te layers, we may assume that the nesting
mechanism operates in each individual Te-Nb-Te sandwich
independently. As a consequence one might argue that the resulting
CDW adjusts its phase in each sandwich, so as to minimize the
repulsive inter-sandwich interaction and to maximize the
attractive intra-sandwich energy. Inspection of Fig.
\ref{fig:struct1} shows, that the Te atoms closest to the Nb
layer, type $c$, falls approximately in between the two Te atoms,
type $a$ and $b$, of the next sandwich which are further away from
the Nb atoms. This maximizes the distance between individual Te
'anions' of successive Te-Nb-Te sandwiches  and increases the
overlap with the Nb 'cations'. In this framework, the tripling of
the unit cell perpendicular to the layers is a consequence of the
in-plane $(3\times 1)$ reconstruction associated with feature 1a,
which in turn is triggered by Fermi surface nesting. This would
also explain, why we do not observe a $(3\times 1\times 2)$
reconstruction associated with feature 1b, since such a
configuration does not minimize the inter-sandwich interaction.\\
A triple-axis distortion as in TaS$_2$ \cite{Fazekas79,Fazekas80},
in which surrounding metal atoms are shifted radially and within
the plane towards a central metal atom to form a contracted star
(see \cite{Aebi01} for a sketch,\cite{Bovet03,Bovet04}), is
reserved to clusters of $6n+1$ metal atoms, where $n$ is the
number of shells surrounding the central atom. This leads to a
very precise condition on the nesting vector
$\mathbf{q}=1/\sqrt{6n+1} \mathbf{a^*}$. Since the Fermi surface
of NbTe$_2$ exhibits dominant nesting at
$\mathbf{q}=\frac{1}{3}\mathbf{a^*}$, such a scenario is excluded,
and the crystal locally selects one of the directions, which leads
to the chain like $(3\times1)$ distortion and the breakup into
domains.
\\
A second maximum (feature 2) at $q=0.19 a^*$ along $\Gamma$K might
account for the star-like $(\sqrt{19}\times\sqrt{19})$ CDW phase,
where three shells ($n=3$), each containing  six metal atoms, are
shifted towards a central atom, although it is slightly displaced
from the ideal value $q=1/\sqrt{19}a^*=0.23 a^*$. This nesting
vector leads to an incommensurate phase, which was observed after
cooling of heat-pulsed crystals to a temperature just above room
temperature \cite{Wilson78}. Our $q=0.19 a^*$ agrees with Wilson's
proposal for nesting across the M and M' points. The absence of an
out-of-plane component of this nesting vector in our calculation
is confirmed by experiment \cite{Wilson78}. Upon cooling to room
temperature, the $(\sqrt{19}\times\sqrt{19})$  CDW rotates away
from $\Gamma$K by $6.6^o$ \cite{Wilson78} to become commensurate
with the parent lattice. Such a second-order
incommensurate-to-commensurate (lock-in) phase transition has been
modelled theoretically via a Landau free energy expansion
\cite{McMillan76}, describing the competition of the terms that
determine the individual periodicities and the term that promotes
commensurability via gap formation. A similar scenario is followed
by the $(\sqrt{13}\times\sqrt{13})$ CDW of \textit{1T}-TaS$_2$,
where in contrast to NbTe$_2$ the nesting vector points along
$\Gamma$M.

\subsection{Phonon bandstructure}
\begin{figure}
\includegraphics{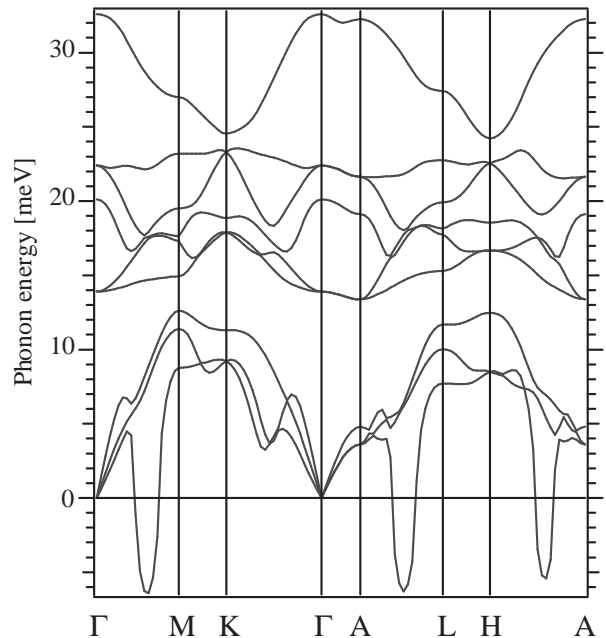}%
\caption{\label{fig:pBS} Phonon dispersion of \textit{1T}-NbTe$_2$
along high symmetry lines obtained by the response function
capabilities of ABINIT. Frequencies below $0$ meV are imaginary.
All free degrees within the trigonal P$\bar3$m1 space group were
relaxed using the Broyden-Fletcher-Goldfarb-Shanno minimization
scheme as implemented in the ABINIT code. The optimized LDA
lattice parameters underestimate the values derived from Brown's
experimental values \cite{Brown66} for the trigonal structure by
less then 1 \%, following the general trend observed for LDA
results \cite{Vanderbilt94}. The LDA equilibrium value for
$z_{red}=0.274$ is also in good agreement with the derived
averaged value $z_{red}=0.277$ (see Appendix \ref{app}). }
\end{figure}
 The occurance of a maximum in the electron susceptibility
alone does not explain the distortion to the monoclinic structure.
The presence of a perturbation with the corresponding $\mathbf{q}$
vector is necessary. In the one-dimensional Peierls scenario this
potential is provided by a soft phonon mode.\\
The DFPT phonon bandstructure for the relaxed trigonal NbTe$_2$
structure obtained by diagonalization of the dynamical matrix
along high symmetry lines is shown in Fig. \ref{fig:pBS}. The
lowest lying acoustic branch exhibits imaginary frequencies. DFPT
contains the implicit assumption that phonons are simple harmonic
modes. Soft modes are by definition anharmonic and their frequency
goes to zero. Zero frequency implies that the lattice structure is
unstable and will transform, typically, to a lower symmetry phase.
In the extreme case, electronic structure calculations may give an
imaginary phonon frequency indicating that the ideal structure is
unstable \cite{Ackland00}. The phonon frequencies are the square
roots of the eigenvalues of the dynamical matrix. Imaginary
frequencies correspond to negative eigenvalues of the dynamical
matrix. A negative entry in the diagonalized dynamical matrix
contributes a negative energy to the total Hamiltonian, indicating
that the expansion was not carried out around the equilibrium
configuration. Thus there exists an energetically more favorable
configuration. At high temperature, the lattice has sufficient
energy to overcome the energy barrier between two or more
symmetry-related variants of the low temperature structure such
that the average observed structure has higher symmetry. In such
cases the ideal structure is stabilized by high temperature and
will undergo a phase transition on cooling, to a low temperature
phase whose symmetry differs by the symmetry of the imaginary
mode.\\
The most unstable modes in Fig. \ref{fig:pBS} occur along
$\mathbf{q}=(1/3,0,q_z)a^*$. This strongly supports the Fermi
surface nesting scenario for NbTe$_2$. Furthermore, from an
analysis of the eigenvectors of the dynamical matrix, the
distorted structure may be qualitatively constructed. The Nb atoms
oscillate predominantly within the basal plane, the Te atoms have
a dominant out of plane component, which is of opposite sign for
the two inequivalent Te atoms.\\
The Raman spectrum of the monoclinic NbTe$_2$ has been measured by
Erdogan and Kirby \cite{Erdogan89} at T=80 K, 300 K, and 420 K. No
phase transition was observed in this temperature range. They
identified 11 peaks in their spectra (see Tab. \ref{tab:table3}),
instead of the two Raman active modes predicted by group theory
for the undistorted compound. Thus unlike in ARPES measurements,
where the unreconstructed (1$\times$1) structure dominates, phonon
bands get backfolded.
\\
A comparison between the experimental Raman peaks and the LDA
results is shown in Tab. \ref{tab:table3}. From a strictly two
dimensional point of view neglecting interlayer effects,
$\mathbf{q}=\frac{1}{3}\mathbf{a^*}$ modes are expected to lie at
the new Brillouin zone center. For comparison with the
experimental data, these modes are included in Tab.
\ref{tab:table3}.
 A low intensity mode
was measured at 31.5 meV and may be identified with the
$\mathbf{q}=0$ mode at 32.6 meV. The low intensity appears to be
reminiscent of its IR character in the undistorted structure. A
second peak is found in the experimental spectrum at 27.2 meV and
may correspond to the backfolded 28.3 meV mode at
$\mathbf{q}=\frac{1}{3}\mathbf{a^*}$.  Nine experimental peaks are
found in the range between 6.9 meV to 21 meV and correspond to the
backfolded acoustic and low-lying optical bands. Including
backfolded LDA modes from $\mathbf{q}=\frac{1}{3}\mathbf{a^*}$,
\textit{ab initio} results indicate the presence of a maximum of
10 modes in the range between 8.8 and 22.4 meV. An additional,
even lower lying acoustic mode is expected due to the
stabilization of the unstable high symmetry mode within the
monoclinic structure. With a rms of relative deviation of 5.8\%
between experiment and LDA, the quantitative agreement can be
qualified as fairly good \cite{Mikami03}. However, not all the LDA
modes are expected to be Raman active. Experimental data for the
IR modes are not available in the literature.
\begin{table}
\caption{\label{tab:table3}Experimental \cite{Erdogan89} and
theoretical Raman modes in meV. Experimental resolution 0.5 meV.
The theoretical optical modes at the Brillouin zone center are
labelled by their corresponding irreducible representation
obtained from a symmetry analysis carried out by ABINIT. Only the
even modes (subscript $g$) are Raman active. The subscript $u$
labels IR active odd modes. For comparison with experiment we also
list the LDA energies of the modes at
$\mathbf{q}=\frac{1}{3}\mathbf{a}*$ which are expected to lie on
the new Brillouin zone center of the distorted structure. }
\begin{ruledtabular}
\begin{tabular}{ccc}
Experiment \cite{Erdogan89}&LDA &backfolded LDA\\
&$\mathbf{q}=0$&$\mathbf{q}=(0.34,0,0)a^*$\\ \hline
 6.9 &  & 6.4\footnotemark[1] \\
 8.7 &  & 8.8 \\
10.4 &  & 9.7 \\
13.0  & 13.9 ($E_g$) &  \\
13.6  &  & 14.7 \\
16.2  &  & 17.5,17.6 \\
18.5  &  & 17.9 \\
19.6  & 20.1 ($A_{1g}$) &  \\
21.0  & 22.4 ($E_u$) & 22.2 \\
27.2  &  & 28.3  \\
31.5  & 32.6 ($A_{1u}$) &  \\
\end{tabular}
\end{ruledtabular}
 \footnotetext[1]{Linearly interpolated between
4.5 meV at $\mathbf{q}=0.2\mathbf{a^*}$ and 8.8 meV at
$\mathbf{q}=0.5\mathbf{a^*}$.}
\end{table}

\section{Comparison with $\mathbf{\mathrm{TaTe}_2}$}

In this article we have concentrated on NbTe$_2$. We want to
stress, however, that most of our conclusions appear to be valid
as well for the isostructural TaTe$_2$. Our experimental LEED and
ARPES data of TaTe$_2$ show a very similar behavior. Furthermore
the electronic bandstructure for trigonal TaTe$_2$ differs only
slightly from the one obtained for NbTe$_2$. We thus conclude that
the nesting behavior of the Fermi surface for TaTe$_2$ is the same
as for NbTe$_2$. However, TaTe$_2$ in contrast to NbTe$_2$ does
not become superconducting at low temperature.\\
Differences between the two compounds are expected in the
vibrational dynamics, since the phonon frequencies scale as the
inverse square root of the mass of the oscillating atoms. The
acoustic branches of the  phonon bandstructure for trigonal
TaTe$_2$ however exhibit a very similar topology as for NbTe$_2$
with unstable modes along $\mathbf{q}=(1/3,0,q_z)a^*$ in the
lowest lying branch. In contrast the bands of the optical manifold
are drastically rearranged. The energy at $\Gamma$ of the highest
lying $A_{1u}$ mode is reduced to 27.6 meV for TaTe$_2$ (32.6 meV
for NbTe$_2$). The $E_u$ mode changes from 22.4 meV for NbTe$_2$
to 19.0 meV for TaTe$_2$ and comes to lie between the two even
modes $E_g$ and $A_{1g}$, whose energies at $\Gamma$ fall within
less than 1 meV onto the energies of the NbTe$_2$ branches. It is
interesting to note that the odd (subscript $u$) IR modes include
Nb/Ta atom motion, whereas for the even (subscript $g$) Raman
modes the Nb/Ta atoms are at rest. This correlates well with the
rescaling of the odd branches and the invariance of the even
branches upon substitution of Nb by the heavier Ta atoms. These
differences in the vibrational spectrum of the two compounds might
possibly explain why TaTe$_2$ is not superconducting, whereas
NbTe$_2$ becomes superconducting below T=0.5-0.74 K.

\section{Summary and Conclusion}

We investigated the origins of the CDW phases of NbTe$_2$. LEED
experiments revealed the presence of three coexisting domains
exhibiting a $(3\times1)$ superstructure, consistent with the
structure derived from X-ray diffraction data, TEM and STM images.
We carried out a detailed \textit{ab initio} study of the nesting
properties of the Fermi surface of the undistorted compound and
found a singularity in the RPA susceptibility at
$\mathbf{q}=\frac{1}{3}\mathbf{a^*}$. In order to consolidate the
absence of an out-of-plane component of this theoretical nesting
vector with the actual $(3\times 1\times 3)$ structure, we suggest
that the CDW  within each individual Te-Nb-Te sandwich adjusts its
phase, so as to minimize the repulsive inter-sandwich interaction
(by maximizing the distance between Te 'anions' of successive
sandwiches) and to maximize the intra-sandwich interaction (by
maximizing the overlap of the Te orbitals with their neighboring
Nb orbitals). A second peak at $q\approx\frac{1}{\sqrt{19}}a^*$
along $\Gamma$K accounts for the $(\sqrt{19}\times\sqrt{19})$ CDW
phase observed by TEM on heat pulsed crystals. \textit{Ab initio}
phonon calculations and a soft
mode analysis support the Fermi surface nesting scenario. \\
Using angle-resolved photoemission in the Fermi surface mapping
mode at room temperature and at T$<$20 K, we found no
quasiparticle crossings in the (3$\times$1) CDW phase of NbTe$_2$.
No phase transition was observed within this temperature range.
Localized gaps at the Fermi level expected for the nesting
scenario could not be observed, since the photoelectron signal is
a superposition of three domains. Instead our ARPES spectra
indicate a pseudogap-like signature over the entire sampled
portion of the Brillouin zone, which can not be understood by
considering only the Peierls scenario and is not reproduced by
DFT. The angular distribution of the spectral weight observed at
the Fermi level is dominated by the residual $(1\times1)$ symmetry
and resembles the metallic DFT Fermi surface of the undistorted
compound. STS spectra indicate that the unoccupied bands are only
weakly affected by the distortion towards the monoclinic
structure, whereas most spectral weight of the Nb $d_{z^2}$ band
is transferred to states below the Fermi energy. The presence of
polarons within the Peierls distorted state possibly accounts for
the absence of any apparent quasiparticle crossing and the
anomalously broad features observed in the ARPES spectra.

\section{Acknowledgments}

We gratefully acknowledge the help of Mirko B\"{o}decker, Marc
Bovet, Oliver Gallus, Christoph Neururer and Thorsten Pillo.
Skillful technical assistance was provided by our workshop and
electric engineering team. This project has in parts been funded
by the Fonds National Suisse pour la Recherche Scientifique.

\appendix*
\section{\label{app}}

The structure derived by Brown \cite{Brown66} is given in Tab.
\ref{tab:table1}. As a starting point for theoretical calculations
within the undistorted trigonal structure, the average values in
table \ref{tab:table2} have been used. The monoclinic cell
parameters, marked with a $m$ subscript are related to the
undistorted cell parameters, without subscript, by
\begin{equation}
a_{m} \approx 3\sqrt{3}a, \; b_{m} \approx a, \; c_{m} \approx
\frac{c}{\sin{\beta}} \label{Mono1Tconversion}
\end{equation}
 For $a$ the value of $b_m$ has been chosen. The
value of $c$ has been obtained via equation
\ref{Mono1Tconversion}. The average value of the $z$ coordinate
for Te was estimated by averaging $z_{red}$ of  Te$_1$, Te$_2$,
Te$_3$ in the monoclinic structure taking into account a small
offset induced by the differences in $z_{red}$ between Nb$_1$ and
Nb$_2$ as well as the multiplicity $n$ of each atom.
\begin{eqnarray}
\bar{z}_{red}(Te)&=&\frac{1}{N(Te)}\sum_{i=1}^3 n(Te_i)
z_{red}(Te_i)\\&&+\Big(1-\frac{1}{N(Nb)}\sum_{i=1}^2 n(Nb_i)
z_{red}(Nb_i)\Big)\nonumber
\end{eqnarray}
where $N(Nb)=6$ and $N(Te)=12$ is the total number of Nb and Te
atoms per unit cell respectively and $z_{red}(Nb_1)=1$.

\begin{table}[h!]
\caption{\label{tab:table1}NbTe$_2$ structure data: $a_{m}=19.39$
\AA, $b_{m}=3.642$ \AA, $c_{m}=9.375$ \AA, $\beta=134.58^o$, space
group 12 (C2/m) \cite{Brown66}.}
\begin{ruledtabular}
\begin{tabular}{lcccr}
Atoms&Point set&x$_{red}$&y$_{red}$&z$_{red}$\\
\hline
Nb$_1$ & 2a & 0.0000 & 0.000 & 0.0000\\
Nb$_2$ & 4i & 0.6397 & 0.000 & 0.9882\\
Te$_1$ & 4i & 0.6497 & 0.000 & 0.2898\\
Te$_2$ & 4i & 0.2970 & 0.000 & 0.2148\\
Te$_3$ & 4i & 0.9961 & 0.000 & 0.3020\\
\end{tabular}
\end{ruledtabular}
\end{table}
\begin{table}[h!]
\caption{\label{tab:table2}Averaged trigonal \textit{1T}-NbTe$_2$
structure data: $a=b=b_{m}=3.642$ \AA, $c=6.678$ \AA, space group
164 (P$\bar3$m1). }
\begin{ruledtabular}
\begin{tabular}{lcccr}
Atoms&Point set&x$_{red}$&y$_{red}$&z$_{red}$\\
\hline
Nb & 1a & 0.0 & 0.0 & 0.0\\
Te & 2d & 1/3 & 2/3 & 0.2767\\
\end{tabular}
\end{ruledtabular}
\end{table}

\bibliography{NbTe2}

\begin{thebibliography}{55}
\expandafter\ifx\csname natexlab\endcsname\relax\def\natexlab#1{#1}\fi
\expandafter\ifx\csname bibnamefont\endcsname\relax
  \def\bibnamefont#1{#1}\fi
\expandafter\ifx\csname bibfnamefont\endcsname\relax
  \def\bibfnamefont#1{#1}\fi
\expandafter\ifx\csname citenamefont\endcsname\relax
  \def\citenamefont#1{#1}\fi
\expandafter\ifx\csname url\endcsname\relax
  \def\url#1{\texttt{#1}}\fi
\expandafter\ifx\csname urlprefix\endcsname\relax\def\urlprefix{URL }\fi
\providecommand{\bibinfo}[2]{#2}
\providecommand{\eprint}[2][]{\url{#2}}

\bibitem[{\citenamefont{Peierls}(1955)}]{Peierls55}
\bibinfo{author}{\bibfnamefont{R.}~\bibnamefont{Peierls}},
  \emph{\bibinfo{title}{Quantum Theory of Solids}}
  (\bibinfo{publisher}{Clarendon}, \bibinfo{year}{1955}).

\bibitem[{\citenamefont{Kohn}(1959)}]{Kohn59}
\bibinfo{author}{\bibfnamefont{W.}~\bibnamefont{Kohn}}, \bibinfo{journal}{Phys.
  Rev. Lett.} \textbf{\bibinfo{volume}{2}}, \bibinfo{pages}{393}
  (\bibinfo{year}{1959}).

\bibitem[{\citenamefont{Brown}(1966)}]{Brown66}
\bibinfo{author}{\bibfnamefont{B.}~\bibnamefont{Brown}}, \bibinfo{journal}{Acta
  Cryst.} \textbf{\bibinfo{volume}{20}}, \bibinfo{pages}{264}
  (\bibinfo{year}{1966}).

\bibitem[{\citenamefont{van Landuyt et~al.}(1974)\citenamefont{van Landuyt, van
  Tendeloo, and Amelinckx}}]{Landuyt74}
\bibinfo{author}{\bibfnamefont{J.}~\bibnamefont{van Landuyt}},
  \bibinfo{author}{\bibfnamefont{G.}~\bibnamefont{van Tendeloo}},
  \bibnamefont{and}
  \bibinfo{author}{\bibfnamefont{S.}~\bibnamefont{Amelinckx}},
  \bibinfo{journal}{Phys. Stat. Sol. (a)} \textbf{\bibinfo{volume}{26}},
  \bibinfo{pages}{585} (\bibinfo{year}{1974}).

\bibitem[{\citenamefont{van Landuyt et~al.}(1975)\citenamefont{van Landuyt, van
  Tendeloo, and Amelinckx}}]{Landuyt75}
\bibinfo{author}{\bibfnamefont{J.}~\bibnamefont{van Landuyt}},
  \bibinfo{author}{\bibfnamefont{G.}~\bibnamefont{van Tendeloo}},
  \bibnamefont{and}
  \bibinfo{author}{\bibfnamefont{S.}~\bibnamefont{Amelinckx}},
  \bibinfo{journal}{Phys. Stat. Sol. (a)} \textbf{\bibinfo{volume}{29}},
  \bibinfo{pages}{K11} (\bibinfo{year}{1975}).

\bibitem[{\citenamefont{Wilson}(1978)}]{Wilson78}
\bibinfo{author}{\bibfnamefont{J.}~\bibnamefont{Wilson}},
  \bibinfo{journal}{Phys. Rev. B} \textbf{\bibinfo{volume}{17}},
  \bibinfo{pages}{3880} (\bibinfo{year}{1978}).

\bibitem[{\citenamefont{Wilson and Yoffe}(1969)}]{Wilson69}
\bibinfo{author}{\bibfnamefont{J.}~\bibnamefont{Wilson}} \bibnamefont{and}
  \bibinfo{author}{\bibfnamefont{A.}~\bibnamefont{Yoffe}},
  \bibinfo{journal}{Adv. Phys.} \textbf{\bibinfo{volume}{18}},
  \bibinfo{pages}{193} (\bibinfo{year}{1969}).

\bibitem[{\citenamefont{Brixner}(1962)}]{Brixner62}
\bibinfo{author}{\bibfnamefont{L.}~\bibnamefont{Brixner}}, \bibinfo{journal}{J.
  Inorg. Nucl. Chem.} \textbf{\bibinfo{volume}{24}}, \bibinfo{pages}{2257}
  (\bibinfo{year}{1962}).

\bibitem[{\citenamefont{Nagata et~al.}(1993)\citenamefont{Nagata, Abe, Ebisu,
  Ishihara, and Tsutsumi}}]{Nagata93}
\bibinfo{author}{\bibfnamefont{S.}~\bibnamefont{Nagata}},
  \bibinfo{author}{\bibfnamefont{T.}~\bibnamefont{Abe}},
  \bibinfo{author}{\bibfnamefont{S.}~\bibnamefont{Ebisu}},
  \bibinfo{author}{\bibfnamefont{Y.}~\bibnamefont{Ishihara}}, \bibnamefont{and}
  \bibinfo{author}{\bibfnamefont{K.}~\bibnamefont{Tsutsumi}},
  \bibinfo{journal}{J. Phys. Chem. Solids} \textbf{\bibinfo{volume}{54}},
  \bibinfo{pages}{895} (\bibinfo{year}{1993}).

\bibitem[{\citenamefont{Vernes et~al.}(1998)\citenamefont{Vernes, Ebert,
  Bensch, Heid, and Naether}}]{Vernes98}
\bibinfo{author}{\bibfnamefont{A.}~\bibnamefont{Vernes}},
  \bibinfo{author}{\bibfnamefont{H.}~\bibnamefont{Ebert}},
  \bibinfo{author}{\bibfnamefont{W.}~\bibnamefont{Bensch}},
  \bibinfo{author}{\bibfnamefont{W.}~\bibnamefont{Heid}}, \bibnamefont{and}
  \bibinfo{author}{\bibfnamefont{C.}~\bibnamefont{Naether}},
  \bibinfo{journal}{J. Phys. Condens. Matter} \textbf{\bibinfo{volume}{10}},
  \bibinfo{pages}{761} (\bibinfo{year}{1998}).

\bibitem[{\citenamefont{van Maaren and Schaeffer}(1967)}]{Maaren67}
\bibinfo{author}{\bibfnamefont{M.}~\bibnamefont{van Maaren}} \bibnamefont{and}
  \bibinfo{author}{\bibfnamefont{G.}~\bibnamefont{Schaeffer}},
  \bibinfo{journal}{Phys. Lett. A} \textbf{\bibinfo{volume}{24}},
  \bibinfo{pages}{645} (\bibinfo{year}{1967}).

\bibitem[{\citenamefont{Kidron}(1967)}]{Kidron67}
\bibinfo{author}{\bibfnamefont{A.}~\bibnamefont{Kidron}},
  \bibinfo{journal}{Phys. Lett. A} \textbf{\bibinfo{volume}{24}},
  \bibinfo{pages}{12} (\bibinfo{year}{1967}).

\bibitem[{\citenamefont{Pillo et~al.}(1999)\citenamefont{Pillo, Hayoz, Berger,
  M.Grioni, Schlapbach, and Aebi}}]{Pillo99b}
\bibinfo{author}{\bibfnamefont{T.}~\bibnamefont{Pillo}},
  \bibinfo{author}{\bibfnamefont{J.}~\bibnamefont{Hayoz}},
  \bibinfo{author}{\bibfnamefont{H.}~\bibnamefont{Berger}},
  \bibinfo{author}{\bibnamefont{M.Grioni}},
  \bibinfo{author}{\bibfnamefont{L.}~\bibnamefont{Schlapbach}},
  \bibnamefont{and} \bibinfo{author}{\bibfnamefont{P.}~\bibnamefont{Aebi}},
  \bibinfo{journal}{Phys. Rev. Lett.} \textbf{\bibinfo{volume}{83}},
  \bibinfo{pages}{3494} (\bibinfo{year}{1999}).

\bibitem[{\citenamefont{Pillo et~al.}(2001)\citenamefont{Pillo, Hayoz,
  Naumovi\'{c}, Berger, Perfetti, Gavioli, Taleb-Ibrahimi, Schapbach, and
  Aebi}}]{Pillo01}
\bibinfo{author}{\bibfnamefont{T.}~\bibnamefont{Pillo}},
  \bibinfo{author}{\bibfnamefont{J.}~\bibnamefont{Hayoz}},
  \bibinfo{author}{\bibfnamefont{D.}~\bibnamefont{Naumovi\'{c}}},
  \bibinfo{author}{\bibfnamefont{H.}~\bibnamefont{Berger}},
  \bibinfo{author}{\bibfnamefont{L.}~\bibnamefont{Perfetti}},
  \bibinfo{author}{\bibfnamefont{L.}~\bibnamefont{Gavioli}},
  \bibinfo{author}{\bibfnamefont{A.}~\bibnamefont{Taleb-Ibrahimi}},
  \bibinfo{author}{\bibfnamefont{L.}~\bibnamefont{Schapbach}},
  \bibnamefont{and} \bibinfo{author}{\bibfnamefont{P.}~\bibnamefont{Aebi}},
  \bibinfo{journal}{Phys. Rev. B} \textbf{\bibinfo{volume}{64}},
  \bibinfo{pages}{2001} (\bibinfo{year}{2001}).

\bibitem[{\citenamefont{Pillo et~al.}(2002)\citenamefont{Pillo, Hayoz, Berger,
  Fasel, Schlapbach, and Aebi}}]{Pillo02}
\bibinfo{author}{\bibfnamefont{T.}~\bibnamefont{Pillo}},
  \bibinfo{author}{\bibfnamefont{J.}~\bibnamefont{Hayoz}},
  \bibinfo{author}{\bibfnamefont{H.}~\bibnamefont{Berger}},
  \bibinfo{author}{\bibfnamefont{R.}~\bibnamefont{Fasel}},
  \bibinfo{author}{\bibfnamefont{L.}~\bibnamefont{Schlapbach}},
  \bibnamefont{and} \bibinfo{author}{\bibfnamefont{P.}~\bibnamefont{Aebi}},
  \bibinfo{journal}{Phys. Rev. B} \textbf{\bibinfo{volume}{62}},
  \bibinfo{pages}{4277} (\bibinfo{year}{2002}).

\bibitem[{\citenamefont{Aebi et~al.}(2001)\citenamefont{Aebi, Pillo, Berger,
  and L\'evy}}]{Aebi01}
\bibinfo{author}{\bibfnamefont{P.}~\bibnamefont{Aebi}},
  \bibinfo{author}{\bibfnamefont{T.}~\bibnamefont{Pillo}},
  \bibinfo{author}{\bibfnamefont{H.}~\bibnamefont{Berger}}, \bibnamefont{and}
  \bibinfo{author}{\bibfnamefont{F.}~\bibnamefont{L\'evy}},
  \bibinfo{journal}{J. Electron Spectrosc. Relat. Phenom.}
  \textbf{\bibinfo{volume}{117}}, \bibinfo{pages}{433} (\bibinfo{year}{2001}).

\bibitem[{\citenamefont{Bovet et~al.}(2003)\citenamefont{Bovet, van Smaalen,
  Berger, Gaal, Forro, Schlapbach, and P.Aebi}}]{Bovet03}
\bibinfo{author}{\bibfnamefont{M.}~\bibnamefont{Bovet}},
  \bibinfo{author}{\bibfnamefont{S.}~\bibnamefont{van Smaalen}},
  \bibinfo{author}{\bibfnamefont{H.}~\bibnamefont{Berger}},
  \bibinfo{author}{\bibfnamefont{R.}~\bibnamefont{Gaal}},
  \bibinfo{author}{\bibfnamefont{L.}~\bibnamefont{Forro}},
  \bibinfo{author}{\bibfnamefont{L.}~\bibnamefont{Schlapbach}},
  \bibnamefont{and} \bibinfo{author}{\bibnamefont{P.Aebi}},
  \bibinfo{journal}{Phys. Rev. B} \textbf{\bibinfo{volume}{67}},
  \bibinfo{pages}{125105} (\bibinfo{year}{2003}).

\bibitem[{\citenamefont{Bovet et~al.}(2004)\citenamefont{Bovet, Popovi\'{c},
  Clerc, Koitzsch, Probst, Bucher, Berger, Naumovi\'{c}, and Aebi}}]{Bovet04}
\bibinfo{author}{\bibfnamefont{M.}~\bibnamefont{Bovet}},
  \bibinfo{author}{\bibfnamefont{D.}~\bibnamefont{Popovi\'{c}}},
  \bibinfo{author}{\bibfnamefont{F.}~\bibnamefont{Clerc}},
  \bibinfo{author}{\bibfnamefont{C.}~\bibnamefont{Koitzsch}},
  \bibinfo{author}{\bibfnamefont{U.}~\bibnamefont{Probst}},
  \bibinfo{author}{\bibfnamefont{E.}~\bibnamefont{Bucher}},
  \bibinfo{author}{\bibfnamefont{H.}~\bibnamefont{Berger}},
  \bibinfo{author}{\bibfnamefont{D.}~\bibnamefont{Naumovi\'{c}}},
  \bibnamefont{and} \bibinfo{author}{\bibfnamefont{P.}~\bibnamefont{Aebi}},
  \bibinfo{journal}{Phys. Rev. B} \textbf{\bibinfo{volume}{69}},
  \bibinfo{pages}{125117} (\bibinfo{year}{2004}).

\bibitem[{\citenamefont{Clerc et~al.}(2004{\natexlab{a}})\citenamefont{Clerc,
  Bovet, Berger, Despont, Koitzsch, and Aebi}}]{Clerc04}
\bibinfo{author}{\bibfnamefont{F.}~\bibnamefont{Clerc}},
  \bibinfo{author}{\bibfnamefont{M.}~\bibnamefont{Bovet}},
  \bibinfo{author}{\bibfnamefont{H.}~\bibnamefont{Berger}},
  \bibinfo{author}{\bibfnamefont{L.}~\bibnamefont{Despont}},
  \bibinfo{author}{\bibfnamefont{C.}~\bibnamefont{Koitzsch}}, \bibnamefont{and}
  \bibinfo{author}{\bibfnamefont{P.}~\bibnamefont{Aebi}},
  \bibinfo{journal}{Physica B} \textbf{\bibinfo{volume}{351}},
  \bibinfo{pages}{245} (\bibinfo{year}{2004}{\natexlab{a}}).

\bibitem[{\citenamefont{Clerc et~al.}(2004{\natexlab{b}})\citenamefont{Clerc,
  Bovet, Berger, Despont, Gallus, Patthey, Shi, Krempasky, Garnier, and
  P.Aebi}}]{Clerc04b}
\bibinfo{author}{\bibfnamefont{F.}~\bibnamefont{Clerc}},
  \bibinfo{author}{\bibfnamefont{M.}~\bibnamefont{Bovet}},
  \bibinfo{author}{\bibfnamefont{H.}~\bibnamefont{Berger}},
  \bibinfo{author}{\bibfnamefont{L.}~\bibnamefont{Despont}},
  \bibinfo{author}{\bibfnamefont{O.}~\bibnamefont{Gallus}},
  \bibinfo{author}{\bibfnamefont{L.}~\bibnamefont{Patthey}},
  \bibinfo{author}{\bibfnamefont{M.}~\bibnamefont{Shi}},
  \bibinfo{author}{\bibfnamefont{J.}~\bibnamefont{Krempasky}},
  \bibinfo{author}{\bibfnamefont{M.}~\bibnamefont{Garnier}}, \bibnamefont{and}
  \bibinfo{author}{\bibnamefont{P.Aebi}}, \bibinfo{journal}{J. Phys. Cond.
  Mat.} \textbf{\bibinfo{volume}{16}}, \bibinfo{pages}{3271}
  (\bibinfo{year}{2004}{\natexlab{b}}).

\bibitem[{\citenamefont{Perfetti et~al.}(2005)\citenamefont{Perfetti, Gloor,
  Mila, Berger, and Grioni}}]{Perfetti05}
\bibinfo{author}{\bibfnamefont{L.}~\bibnamefont{Perfetti}},
  \bibinfo{author}{\bibfnamefont{T.}~\bibnamefont{Gloor}},
  \bibinfo{author}{\bibfnamefont{F.}~\bibnamefont{Mila}},
  \bibinfo{author}{\bibfnamefont{H.}~\bibnamefont{Berger}}, \bibnamefont{and}
  \bibinfo{author}{\bibfnamefont{M.}~\bibnamefont{Grioni}},
  \bibinfo{journal}{Phys. Rev. B} \textbf{\bibinfo{volume}{71}},
  \bibinfo{pages}{153101} (\bibinfo{year}{2005}).

\bibitem[{\citenamefont{Horiba et~al.}(2002)\citenamefont{Horiba, Ono, Oh,
  Kihara, Nakazono, Oshima, Shiino, Yeom, Kakizaki, and Aiura}}]{Horiba02}
\bibinfo{author}{\bibfnamefont{K.}~\bibnamefont{Horiba}},
  \bibinfo{author}{\bibfnamefont{K.}~\bibnamefont{Ono}},
  \bibinfo{author}{\bibfnamefont{J.}~\bibnamefont{Oh}},
  \bibinfo{author}{\bibfnamefont{T.}~\bibnamefont{Kihara}},
  \bibinfo{author}{\bibfnamefont{S.}~\bibnamefont{Nakazono}},
  \bibinfo{author}{\bibfnamefont{M.}~\bibnamefont{Oshima}},
  \bibinfo{author}{\bibfnamefont{O.}~\bibnamefont{Shiino}},
  \bibinfo{author}{\bibfnamefont{H.}~\bibnamefont{Yeom}},
  \bibinfo{author}{\bibfnamefont{A.}~\bibnamefont{Kakizaki}}, \bibnamefont{and}
  \bibinfo{author}{\bibfnamefont{Y.}~\bibnamefont{Aiura}},
  \bibinfo{journal}{Phys. Rev. B} \textbf{\bibinfo{volume}{66}},
  \bibinfo{pages}{073106} (\bibinfo{year}{2002}).

\bibitem[{\citenamefont{Perfetti et~al.}(2003)\citenamefont{Perfetti, Georges,
  Florens, Biermann, Mitrovic, Berger, Tomm, H$\mathrm{\ddot{o}}$chst, and
  Grioni}}]{Perfetti03}
\bibinfo{author}{\bibfnamefont{L.}~\bibnamefont{Perfetti}},
  \bibinfo{author}{\bibfnamefont{A.}~\bibnamefont{Georges}},
  \bibinfo{author}{\bibfnamefont{S.}~\bibnamefont{Florens}},
  \bibinfo{author}{\bibfnamefont{S.}~\bibnamefont{Biermann}},
  \bibinfo{author}{\bibfnamefont{S.}~\bibnamefont{Mitrovic}},
  \bibinfo{author}{\bibfnamefont{H.}~\bibnamefont{Berger}},
  \bibinfo{author}{\bibfnamefont{Y.}~\bibnamefont{Tomm}},
  \bibinfo{author}{\bibfnamefont{H.}~\bibnamefont{H$\mathrm{\ddot{o}}$chst}},
  \bibnamefont{and} \bibinfo{author}{\bibfnamefont{M.}~\bibnamefont{Grioni}},
  \bibinfo{journal}{Phys. Rev. Lett.} \textbf{\bibinfo{volume}{90}},
  \bibinfo{pages}{166401} (\bibinfo{year}{2003}).

\bibitem[{\citenamefont{Colonna et~al.}(2005)\citenamefont{Colonna, Ronci,
  Cricenti, Perfetti, Berger, and Grioni}}]{Colonna05}
\bibinfo{author}{\bibfnamefont{S.}~\bibnamefont{Colonna}},
  \bibinfo{author}{\bibfnamefont{R.}~\bibnamefont{Ronci}},
  \bibinfo{author}{\bibfnamefont{A.}~\bibnamefont{Cricenti}},
  \bibinfo{author}{\bibfnamefont{L.}~\bibnamefont{Perfetti}},
  \bibinfo{author}{\bibfnamefont{H.}~\bibnamefont{Berger}}, \bibnamefont{and}
  \bibinfo{author}{\bibfnamefont{M.}~\bibnamefont{Grioni}},
  \bibinfo{journal}{Phys. Rev. Lett.} \textbf{\bibinfo{volume}{94}},
  \bibinfo{pages}{036405} (\bibinfo{year}{2005}).

\bibitem[{\citenamefont{Pillo et~al.}(1998)\citenamefont{Pillo, Patthey,
  Boschung, Hayoz, Aebi, and Schlapbach}}]{Pillo98}
\bibinfo{author}{\bibfnamefont{T.}~\bibnamefont{Pillo}},
  \bibinfo{author}{\bibfnamefont{L.}~\bibnamefont{Patthey}},
  \bibinfo{author}{\bibfnamefont{E.}~\bibnamefont{Boschung}},
  \bibinfo{author}{\bibfnamefont{J.}~\bibnamefont{Hayoz}},
  \bibinfo{author}{\bibfnamefont{P.}~\bibnamefont{Aebi}}, \bibnamefont{and}
  \bibinfo{author}{\bibfnamefont{L.}~\bibnamefont{Schlapbach}},
  \bibinfo{journal}{J. Electron Spectr. Relat. Phenom.}
  \textbf{\bibinfo{volume}{97}}, \bibinfo{pages}{243} (\bibinfo{year}{1998}).

\bibitem[{\citenamefont{Aebi et~al.}(1998)\citenamefont{Aebi, Fasel,
  Naumovi\'c, Hayoz, Pillo, Bovet, Agostino, Patthey, Schlapbach, Gil
  et~al.}}]{Aebi97}
\bibinfo{author}{\bibfnamefont{P.}~\bibnamefont{Aebi}},
  \bibinfo{author}{\bibfnamefont{R.}~\bibnamefont{Fasel}},
  \bibinfo{author}{\bibfnamefont{D.}~\bibnamefont{Naumovi\'c}},
  \bibinfo{author}{\bibfnamefont{J.}~\bibnamefont{Hayoz}},
  \bibinfo{author}{\bibfnamefont{T.}~\bibnamefont{Pillo}},
  \bibinfo{author}{\bibfnamefont{M.}~\bibnamefont{Bovet}},
  \bibinfo{author}{\bibfnamefont{R.}~\bibnamefont{Agostino}},
  \bibinfo{author}{\bibfnamefont{L.}~\bibnamefont{Patthey}},
  \bibinfo{author}{\bibfnamefont{L.}~\bibnamefont{Schlapbach}},
  \bibinfo{author}{\bibfnamefont{F.}~\bibnamefont{Gil}}, \bibnamefont{et~al.},
  \bibinfo{journal}{Surf. Sci.} \textbf{\bibinfo{volume}{402-404}},
  \bibinfo{pages}{614} (\bibinfo{year}{1998}).

\bibitem[{\citenamefont{Fasel and Aebi}(2002)}]{Fasel02}
\bibinfo{author}{\bibfnamefont{R.}~\bibnamefont{Fasel}} \bibnamefont{and}
  \bibinfo{author}{\bibfnamefont{P.}~\bibnamefont{Aebi}},
  \bibinfo{journal}{Chimia} \textbf{\bibinfo{volume}{56}}, \bibinfo{pages}{566}
  (\bibinfo{year}{2002}).

\bibitem[{\citenamefont{Perdew et~al.}(1996)\citenamefont{Perdew, Burke, and
  Ernznerhof}}]{Perdew96}
\bibinfo{author}{\bibfnamefont{J.}~\bibnamefont{Perdew}},
  \bibinfo{author}{\bibfnamefont{S.}~\bibnamefont{Burke}}, \bibnamefont{and}
  \bibinfo{author}{\bibfnamefont{M.}~\bibnamefont{Ernznerhof}},
  \bibinfo{journal}{Phys. Rev. Lett.} \textbf{\bibinfo{volume}{77}},
  \bibinfo{pages}{3865} (\bibinfo{year}{1996}).

\bibitem[{\citenamefont{Blaha et~al.}(2001)\citenamefont{Blaha, Schwarz,
  Madsen, Kvasnicka, and Luitz}}]{wien2k}
\bibinfo{author}{\bibfnamefont{P.}~\bibnamefont{Blaha}},
  \bibinfo{author}{\bibfnamefont{K.}~\bibnamefont{Schwarz}},
  \bibinfo{author}{\bibfnamefont{G.}~\bibnamefont{Madsen}},
  \bibinfo{author}{\bibfnamefont{D.}~\bibnamefont{Kvasnicka}},
  \bibnamefont{and} \bibinfo{author}{\bibfnamefont{J.}~\bibnamefont{Luitz}}
  (\bibinfo{year}{2001}), \eprint{WIEN2k, An Augmented Plane Wave + Local
  Orbitals Program for Calculating Crystal Properties (Karlheinz Schwarz, Tech.
  Univ. Wien, Austria) ISBN 3-9501031-1-2}.

\bibitem[{\citenamefont{Gonze et~al.}(2002)\citenamefont{Gonze, Beuken,
  Caracas, Detraux, Fuchs, Rignanese, Sindic, Verstaete, Zerah, Jollet
  et~al.}}]{Gonze02}
\bibinfo{author}{\bibfnamefont{X.}~\bibnamefont{Gonze}},
  \bibinfo{author}{\bibfnamefont{J.-M.} \bibnamefont{Beuken}},
  \bibinfo{author}{\bibfnamefont{R.}~\bibnamefont{Caracas}},
  \bibinfo{author}{\bibfnamefont{F.}~\bibnamefont{Detraux}},
  \bibinfo{author}{\bibfnamefont{M.}~\bibnamefont{Fuchs}},
  \bibinfo{author}{\bibfnamefont{G.-M.} \bibnamefont{Rignanese}},
  \bibinfo{author}{\bibfnamefont{L.}~\bibnamefont{Sindic}},
  \bibinfo{author}{\bibfnamefont{M.}~\bibnamefont{Verstaete}},
  \bibinfo{author}{\bibfnamefont{G.}~\bibnamefont{Zerah}},
  \bibinfo{author}{\bibfnamefont{F.}~\bibnamefont{Jollet}},
  \bibnamefont{et~al.}, \bibinfo{journal}{Computational Materials Science}
  \textbf{\bibinfo{volume}{25}}, \bibinfo{pages}{478} (\bibinfo{year}{2002}).

\bibitem[{abi()}]{abinit}
\eprint{The ABINIT code is a common project of the Universit\'{e} Catholique de
  Louvain, Corning Incorporated, and other contributors (URL
  http://www.abinit.org)}.

\bibitem[{\citenamefont{Hartwigsen et~al.}(1998)\citenamefont{Hartwigsen,
  Goedecker, and Hutter}}]{Hartwigsen98}
\bibinfo{author}{\bibfnamefont{C.}~\bibnamefont{Hartwigsen}},
  \bibinfo{author}{\bibfnamefont{S.}~\bibnamefont{Goedecker}},
  \bibnamefont{and} \bibinfo{author}{\bibfnamefont{J.}~\bibnamefont{Hutter}},
  \bibinfo{journal}{Phys. Rev. B} \textbf{\bibinfo{volume}{58}},
  \bibinfo{pages}{3641} (\bibinfo{year}{1998}).

\bibitem[{\citenamefont{Ambrosch-Draxl and Sofo}(2004)}]{Ambrosch04}
\bibinfo{author}{\bibfnamefont{C.}~\bibnamefont{Ambrosch-Draxl}}
  \bibnamefont{and} \bibinfo{author}{\bibfnamefont{J.~O.} \bibnamefont{Sofo}}
  (\bibinfo{year}{2004}), \eprint{cond-mat/0402523}.

\bibitem[{\citenamefont{Gonze}(1997)}]{Gonze97}
\bibinfo{author}{\bibfnamefont{X.}~\bibnamefont{Gonze}},
  \bibinfo{journal}{Phys. Rev. B} \textbf{\bibinfo{volume}{55}},
  \bibinfo{pages}{10337} (\bibinfo{year}{1997}).

\bibitem[{\citenamefont{Gonze and Lee}(1997)}]{Gonze97b}
\bibinfo{author}{\bibfnamefont{X.}~\bibnamefont{Gonze}} \bibnamefont{and}
  \bibinfo{author}{\bibfnamefont{C.}~\bibnamefont{Lee}},
  \bibinfo{journal}{Phys. Rev. B} \textbf{\bibinfo{volume}{55}},
  \bibinfo{pages}{10355} (\bibinfo{year}{1997}).

\bibitem[{DFT()}]{DFTdetails}
\bibinfo{note}{In WIEN2k, the APW+lo basis was expanded up to $R_{MT}K_{MAX} =
  7$, where $K_{MAX}$ is the maximum modulus for the reciprocal lattice vector
  and $R_{MT}$ is the radius of the muffin tin sphere. Inside the muffin tin
  sphere, the $l$-expansion of the non-spherical potential and charge density
  is carried out up to $l_{max} = 10$. For the trigonal structure,
  k-integration over the Brillouin zone is performed using $2000$ k-points. For
  the computation of the static susceptibility, integration over the Brillouin
  zone has been carried out on a $81\times81\times21$ k-point grid. It is not
  possible to compute a strictly static susceptibility. An energy window of 1.0
  mRy has been chosen for all calculations. Results obtained using an energy
  window of 0.1 mRy did not differ significantly. For groundstate calculations
  in ABINIT, the wave functions were expanded in a planewave basis with cutoff
  energy of $40$ Hartree. For the trigonal structure, Brillouin zone
  integration is performed on a $8\times 8\times 4$ Monkhorst-Pack k-point
  mesh. For the computation of the phonon bandstructure of the trigonal
  NbTe$_2$ compound, 35 dynamical matrices have been computed resulting from an
  unshifted $8\times 8\times 4$ Monkhorst-Pack Brillouin zone sampling using a
  cutoff energy of $20$ Hartree.}

\bibitem[{\citenamefont{Cukjati et~al.}(2002)\citenamefont{Cukjati, Prodan,
  Jug, van Midden, Hla, Boehm, Boswell, and Bennett}}]{Cukjati02b}
\bibinfo{author}{\bibfnamefont{D.}~\bibnamefont{Cukjati}},
  \bibinfo{author}{\bibfnamefont{A.}~\bibnamefont{Prodan}},
  \bibinfo{author}{\bibfnamefont{N.}~\bibnamefont{Jug}},
  \bibinfo{author}{\bibfnamefont{H.}~\bibnamefont{van Midden}},
  \bibinfo{author}{\bibfnamefont{S.}~\bibnamefont{Hla}},
  \bibinfo{author}{\bibfnamefont{H.}~\bibnamefont{Boehm}},
  \bibinfo{author}{\bibfnamefont{F.}~\bibnamefont{Boswell}}, \bibnamefont{and}
  \bibinfo{author}{\bibfnamefont{J.}~\bibnamefont{Bennett}},
  \bibinfo{journal}{Phys. Stat. Sol. (a)} \textbf{\bibinfo{volume}{193}},
  \bibinfo{pages}{246} (\bibinfo{year}{2002}).

\bibitem[{\citenamefont{Mesot et~al.}(2001)\citenamefont{Mesot, Randeria,
  Norman, Kaminski, Fretwell, Campuzano, Ding, Takeuchi, Sato, Yokoya
  et~al.}}]{Mesot01}
\bibinfo{author}{\bibfnamefont{J.}~\bibnamefont{Mesot}},
  \bibinfo{author}{\bibfnamefont{M.}~\bibnamefont{Randeria}},
  \bibinfo{author}{\bibfnamefont{M.}~\bibnamefont{Norman}},
  \bibinfo{author}{\bibfnamefont{A.}~\bibnamefont{Kaminski}},
  \bibinfo{author}{\bibfnamefont{H.}~\bibnamefont{Fretwell}},
  \bibinfo{author}{\bibfnamefont{J.}~\bibnamefont{Campuzano}},
  \bibinfo{author}{\bibfnamefont{H.}~\bibnamefont{Ding}},
  \bibinfo{author}{\bibfnamefont{T.}~\bibnamefont{Takeuchi}},
  \bibinfo{author}{\bibfnamefont{T.}~\bibnamefont{Sato}},
  \bibinfo{author}{\bibfnamefont{T.}~\bibnamefont{Yokoya}},
  \bibnamefont{et~al.}, \bibinfo{journal}{Phys. Rev. B}
  \textbf{\bibinfo{volume}{63}}, \bibinfo{pages}{224516}
  (\bibinfo{year}{2001}).

\bibitem[{\citenamefont{Damascelli et~al.}(2003)\citenamefont{Damascelli, Shen,
  and Hussain}}]{Damascelli03}
\bibinfo{author}{\bibfnamefont{A.}~\bibnamefont{Damascelli}},
  \bibinfo{author}{\bibfnamefont{Z.-X.} \bibnamefont{Shen}}, \bibnamefont{and}
  \bibinfo{author}{\bibfnamefont{Z.}~\bibnamefont{Hussain}},
  \bibinfo{journal}{Rev. Mod. Phys.} \textbf{\bibinfo{volume}{75}},
  \bibinfo{pages}{473} (\bibinfo{year}{2003}).

\bibitem[{\citenamefont{Voit et~al.}(2000)\citenamefont{Voit, Perfetti, Zwick,
  Berger, Margaritondo, Gr$\mathrm{\ddot{u}}$ner, H$\mathrm{\ddot{o}}$chst, and
  Grioni}}]{Voit00}
\bibinfo{author}{\bibfnamefont{J.}~\bibnamefont{Voit}},
  \bibinfo{author}{\bibfnamefont{L.}~\bibnamefont{Perfetti}},
  \bibinfo{author}{\bibfnamefont{F.}~\bibnamefont{Zwick}},
  \bibinfo{author}{\bibfnamefont{H.}~\bibnamefont{Berger}},
  \bibinfo{author}{\bibfnamefont{G.}~\bibnamefont{Margaritondo}},
  \bibinfo{author}{\bibfnamefont{G.}~\bibnamefont{Gr$\mathrm{\ddot{u}}$ner}},
  \bibinfo{author}{\bibfnamefont{H.}~\bibnamefont{H$\mathrm{\ddot{o}}$chst}},
  \bibnamefont{and} \bibinfo{author}{\bibfnamefont{M.}~\bibnamefont{Grioni}},
  \bibinfo{journal}{Science} \textbf{\bibinfo{volume}{290}},
  \bibinfo{pages}{501} (\bibinfo{year}{2000}).

\bibitem[{\citenamefont{Dessau et~al.}(1999)\citenamefont{Dessau, Saitoh, Park,
  Shen, Villella, Hamada, Moritomo, and Tokura}}]{Dessau99}
\bibinfo{author}{\bibfnamefont{D.}~\bibnamefont{Dessau}},
  \bibinfo{author}{\bibfnamefont{T.}~\bibnamefont{Saitoh}},
  \bibinfo{author}{\bibfnamefont{C.-H.} \bibnamefont{Park}},
  \bibinfo{author}{\bibfnamefont{Z.-X.} \bibnamefont{Shen}},
  \bibinfo{author}{\bibfnamefont{P.}~\bibnamefont{Villella}},
  \bibinfo{author}{\bibfnamefont{N.}~\bibnamefont{Hamada}},
  \bibinfo{author}{\bibfnamefont{Y.}~\bibnamefont{Moritomo}}, \bibnamefont{and}
  \bibinfo{author}{\bibfnamefont{Y.}~\bibnamefont{Tokura}},
  \bibinfo{journal}{J. Superconductivity} \textbf{\bibinfo{volume}{12}},
  \bibinfo{pages}{273} (\bibinfo{year}{1999}).

\bibitem[{\citenamefont{Perfetti et~al.}(2001)\citenamefont{Perfetti, Berger,
  Reginelli, Degiorgi, H$\mathrm{\ddot{o}}$chst, Voit, Margaritondo, and
  Grioni}}]{Perfetti01}
\bibinfo{author}{\bibfnamefont{L.}~\bibnamefont{Perfetti}},
  \bibinfo{author}{\bibfnamefont{H.}~\bibnamefont{Berger}},
  \bibinfo{author}{\bibfnamefont{A.}~\bibnamefont{Reginelli}},
  \bibinfo{author}{\bibfnamefont{L.}~\bibnamefont{Degiorgi}},
  \bibinfo{author}{\bibfnamefont{H.}~\bibnamefont{H$\mathrm{\ddot{o}}$chst}},
  \bibinfo{author}{\bibfnamefont{J.}~\bibnamefont{Voit}},
  \bibinfo{author}{\bibfnamefont{G.}~\bibnamefont{Margaritondo}},
  \bibnamefont{and} \bibinfo{author}{\bibfnamefont{M.}~\bibnamefont{Grioni}},
  \bibinfo{journal}{Phys. Rev. Lett.} \textbf{\bibinfo{volume}{87}},
  \bibinfo{pages}{216404} (\bibinfo{year}{2001}).

\bibitem[{\citenamefont{Perfetti et~al.}(2002)\citenamefont{Perfetti, Mitrovic,
  Margaritondo, Grioni, Forr\'{o}, Degiorgi, and
  H$\mathrm{\ddot{o}}$chst}}]{Perfetti02}
\bibinfo{author}{\bibfnamefont{L.}~\bibnamefont{Perfetti}},
  \bibinfo{author}{\bibfnamefont{S.}~\bibnamefont{Mitrovic}},
  \bibinfo{author}{\bibfnamefont{G.}~\bibnamefont{Margaritondo}},
  \bibinfo{author}{\bibfnamefont{M.}~\bibnamefont{Grioni}},
  \bibinfo{author}{\bibfnamefont{L.}~\bibnamefont{Forr\'{o}}},
  \bibinfo{author}{\bibfnamefont{L.}~\bibnamefont{Degiorgi}}, \bibnamefont{and}
  \bibinfo{author}{\bibfnamefont{H.}~\bibnamefont{H$\mathrm{\ddot{o}}$chst}},
  \bibinfo{journal}{Phys. Rev. B} \textbf{\bibinfo{volume}{66}},
  \bibinfo{pages}{075107} (\bibinfo{year}{2002}).

\bibitem[{\citenamefont{Hohenadler et~al.}(2005)\citenamefont{Hohenadler,
  Neuber, von~der Linden, Wellein, and Fehske}}]{Hohenadler05}
\bibinfo{author}{\bibfnamefont{M.}~\bibnamefont{Hohenadler}},
  \bibinfo{author}{\bibfnamefont{D.}~\bibnamefont{Neuber}},
  \bibinfo{author}{\bibfnamefont{W.}~\bibnamefont{von~der Linden}},
  \bibinfo{author}{\bibfnamefont{G.}~\bibnamefont{Wellein}}, \bibnamefont{and}
  \bibinfo{author}{\bibfnamefont{H.}~\bibnamefont{Fehske}},
  \bibinfo{journal}{Phys. Rev. B} \textbf{\bibinfo{volume}{71}},
  \bibinfo{pages}{245111} (\bibinfo{year}{2005}).

\bibitem[{\citenamefont{Sykora et~al.}(2005)\citenamefont{Sykora,
  H$\mathrm{\ddot{u}}$bsch, Becker, Wellein, and Fehske}}]{Sykora05}
\bibinfo{author}{\bibfnamefont{S.}~\bibnamefont{Sykora}},
  \bibinfo{author}{\bibfnamefont{A.}~\bibnamefont{H$\mathrm{\ddot{u}}$bsch}},
  \bibinfo{author}{\bibfnamefont{K.}~\bibnamefont{Becker}},
  \bibinfo{author}{\bibfnamefont{G.}~\bibnamefont{Wellein}}, \bibnamefont{and}
  \bibinfo{author}{\bibfnamefont{H.}~\bibnamefont{Fehske}},
  \bibinfo{journal}{Phys. Rev. B} \textbf{\bibinfo{volume}{71}},
  \bibinfo{pages}{045112} (\bibinfo{year}{2005}).

\bibitem[{\citenamefont{Kornilovitch}(2002)}]{Kornilovitch02}
\bibinfo{author}{\bibfnamefont{P.}~\bibnamefont{Kornilovitch}},
  \bibinfo{journal}{Europhys. Lett.} \textbf{\bibinfo{volume}{59}},
  \bibinfo{pages}{735} (\bibinfo{year}{2002}).

\bibitem[{\citenamefont{Weitering et~al.}(1996)\citenamefont{Weitering, Shi,
  and Erwin}}]{Weitering96}
\bibinfo{author}{\bibfnamefont{H.}~\bibnamefont{Weitering}},
  \bibinfo{author}{\bibfnamefont{X.}~\bibnamefont{Shi}}, \bibnamefont{and}
  \bibinfo{author}{\bibfnamefont{S.}~\bibnamefont{Erwin}},
  \bibinfo{journal}{Phys. Rev. B} \textbf{\bibinfo{volume}{54}},
  \bibinfo{pages}{10585} (\bibinfo{year}{1996}).

\bibitem[{\citenamefont{Fazekas and Tosatti}(1979)}]{Fazekas79}
\bibinfo{author}{\bibfnamefont{P.}~\bibnamefont{Fazekas}} \bibnamefont{and}
  \bibinfo{author}{\bibfnamefont{E.}~\bibnamefont{Tosatti}},
  \bibinfo{journal}{Phil. Mag. B 39} \textbf{\bibinfo{volume}{39}},
  \bibinfo{pages}{229} (\bibinfo{year}{1979}).

\bibitem[{\citenamefont{Koitzsch et~al.}(2004)\citenamefont{Koitzsch, Hayoz,
  Bovet, Clerc, Despont, Ambrosch-Draxl, and Aebi}}]{Koitzsch04}
\bibinfo{author}{\bibfnamefont{C.}~\bibnamefont{Koitzsch}},
  \bibinfo{author}{\bibfnamefont{J.}~\bibnamefont{Hayoz}},
  \bibinfo{author}{\bibfnamefont{M.}~\bibnamefont{Bovet}},
  \bibinfo{author}{\bibfnamefont{F.}~\bibnamefont{Clerc}},
  \bibinfo{author}{\bibfnamefont{L.}~\bibnamefont{Despont}},
  \bibinfo{author}{\bibfnamefont{C.}~\bibnamefont{Ambrosch-Draxl}},
  \bibnamefont{and} \bibinfo{author}{\bibfnamefont{P.}~\bibnamefont{Aebi}},
  \bibinfo{journal}{Phys. Rev. B} \textbf{\bibinfo{volume}{70}},
  \bibinfo{pages}{165114} (\bibinfo{year}{2004}).

\bibitem[{\citenamefont{Fazekas and Tosatti}(1980)}]{Fazekas80}
\bibinfo{author}{\bibfnamefont{P.}~\bibnamefont{Fazekas}} \bibnamefont{and}
  \bibinfo{author}{\bibfnamefont{E.}~\bibnamefont{Tosatti}},
  \bibinfo{journal}{Physica B} \textbf{\bibinfo{volume}{99}},
  \bibinfo{pages}{183} (\bibinfo{year}{1980}).

\bibitem[{\citenamefont{McMillan}(1976)}]{McMillan76}
\bibinfo{author}{\bibfnamefont{W.}~\bibnamefont{McMillan}},
  \bibinfo{journal}{Phys. Rev. B} \textbf{\bibinfo{volume}{14}},
  \bibinfo{pages}{1496} (\bibinfo{year}{1976}).

\bibitem[{\citenamefont{King-Smith and Vanderbilt}(1994)}]{Vanderbilt94}
\bibinfo{author}{\bibfnamefont{R.}~\bibnamefont{King-Smith}} \bibnamefont{and}
  \bibinfo{author}{\bibfnamefont{D.}~\bibnamefont{Vanderbilt}},
  \bibinfo{journal}{Phys. Rev. B} \textbf{\bibinfo{volume}{49}},
  \bibinfo{pages}{5828} (\bibinfo{year}{1994}).

\bibitem[{\citenamefont{Ackland}(2000)}]{Ackland00}
\bibinfo{author}{\bibfnamefont{G.}~\bibnamefont{Ackland}},
  \bibinfo{journal}{RIKEN review} \textbf{\bibinfo{volume}{29}},
  \bibinfo{pages}{34} (\bibinfo{year}{2000}).

\bibitem[{\citenamefont{Erdogan and Kirby}(1989)}]{Erdogan89}
\bibinfo{author}{\bibfnamefont{H.}~\bibnamefont{Erdogan}} \bibnamefont{and}
  \bibinfo{author}{\bibfnamefont{R.}~\bibnamefont{Kirby}},
  \bibinfo{journal}{Sol. Stat. Commun.} \textbf{\bibinfo{volume}{70}},
  \bibinfo{pages}{713} (\bibinfo{year}{1989}).

\bibitem[{\citenamefont{Mikami et~al.}(2003)\citenamefont{Mikami, Nakamura,
  Itoh, Nakajima, and Shishido}}]{Mikami03}
\bibinfo{author}{\bibfnamefont{M.}~\bibnamefont{Mikami}},
  \bibinfo{author}{\bibfnamefont{S.}~\bibnamefont{Nakamura}},
  \bibinfo{author}{\bibfnamefont{M.}~\bibnamefont{Itoh}},
  \bibinfo{author}{\bibfnamefont{K.}~\bibnamefont{Nakajima}}, \bibnamefont{and}
  \bibinfo{author}{\bibfnamefont{T.}~\bibnamefont{Shishido}},
  \bibinfo{journal}{J. Luminescence} \textbf{\bibinfo{volume}{102-103}},
  \bibinfo{pages}{7} (\bibinfo{year}{2003}).

\end{thebibliography}

\end{document}